%% file: LLM-dense-PRF.tex
\documentclass[sigconf,natbib=true,anonymous=false]{acmart}

\usepackage[T1]{fontenc}
\usepackage{bbold}
\usepackage{marginnote}
\usepackage{subcaption}
\usepackage[show]{chato-notes}
\usepackage{float}
\usepackage{graphicx}
\usepackage{booktabs,siunitx}
\usepackage{bbold}
\usepackage{amsmath}
\usepackage{natbib}
\usepackage{amsfonts}
\usepackage{subcaption}
\usepackage{hhline}
\usepackage{hyperref}
\usepackage{enumerate}
\usepackage{multirow} 
\usepackage{array}
\usepackage{bbm}
\usepackage{adjustbox}
\usepackage[algo2e,ruled,vlined]{algorithm2e}
\usepackage[framemethod=TikZ]{mdframed}
\usepackage{tabularx, makecell, multirow} 
\usepackage{lipsum}
\usepackage{schemata}
\usepackage[bottom]{footmisc}
\usepackage{ulem} % scrikeout text
\usepackage{enumitem}

\DeclareMathSymbol{\mlq}{\mathord}{operators}{``}
\DeclareMathSymbol{\mrq}{\mathord}{operators}{`'}
\AtBeginDocument{%
  }

\setcopyright{acmlicensed}
\copyrightyear{2025}
\acmYear{2025}
\acmDOI{XXXXXXX.XXXXXXX}
\acmConference[Conference acronym 'XX]{Make sure to enter the correct
  conference title from your rights confirmation email}{June 03--05,
  2025}{Woodstock, NY}
\acmISBN{978-1-4503-XXXX-X/2025/06}

\begin{document}
%%
%% The "title" command has an optional parameter,
%% allowing the author to define a "short title" to be used in page headers.
%\title{\sout{Scaling Isn’t Everything: Closing the Gap Between Small and Large Retrievers}
\title{Pseudo Relevance Feedback is Enough to Close the Gap Between Small and Large Dense Retrieval Models}

%%
%% The "author" command and its associated commands are used to define
%% the authors and their affiliations.
%% Of note is the shared affiliation of the first two authors, and the
%% "authornote" and "authornotemark" commands
%% used to denote shared contribution to the research.
\author{Hang Li}
\orcid{0000-0002-5317-7227}
\affiliation{%
  \institution{The University of Queensland}
  \city{Brisbane}
  \country{Australia}
}
\email{hang.li@uq.edu.au}

\author{Xiao Wang}
\orcid{0000-0002-5151-2773}
\affiliation{%
  \institution{University of International Business and Economics}
  \city{Beijing}
  \country{China}}
\email{xiao.wang@uibe.edu.cn}

\author{Bevan Koopman}
\orcid{0000-0001-5577-3391}
\affiliation{%
  \institution{CSIRO \& The University if Queensland}
  \city{Brisbane}
  \country{Australia}}
\email{bevan.koopman@csiro.au}

\author{Guido Zuccon}
\orcid{0000-0003-0271-5563}
\affiliation{%
 \institution{The University of Queensland}
 \city{Brisbane}
 \country{Australia}}
\email{g.zuccon@uq.edu.au}

%%
%% By default, the full list of authors will be used in the page
%% headers. Often, this list is too long, and will overlap
%% other information printed in the page headers. This command allows
%% the author to define a more concise list
%% of authors' names for this purpose.
\renewcommand{\shortauthors}{Hang et al.}

%%
%% The abstract is a short summary of the work to be presented in the
%% article.
\begin{abstract}

Scaling dense retrievers to larger large language model (LLM) backbones has been a dominant strategy for improving their retrieval effectiveness.
However, this has substantial cost implications: larger backbones require more expensive hardware (e.g. GPUs with more memory) and lead to higher indexing and querying costs (latency, energy consumption).
In this paper, we challenge this paradigm by introducing PromptPRF, a feature-based pseudo-relevance feedback (PRF) framework that enables small LLM-based dense retrievers to achieve effectiveness comparable to much larger models.

PromptPRF uses LLMs to extract query-independent, structured and unstructured features (e.g., entities, summaries, chain-of-thought keywords, essay) from top-ranked documents. These features are generated offline and integrated into dense query representations via prompting, enabling efficient retrieval without additional training. 
Unlike prior methods such as GRF, which rely on online, query-specific generation and sparse retrieval, 
PromptPRF decouples feedback generation from query processing and supports dense retrievers in a fully zero-shot setting.

Experiments on TREC DL and BEIR benchmarks demonstrate that PromptPRF consistently improves retrieval effectiveness and offers favourable cost-effectiveness trade-offs. 
We further present ablation studies to understand the role of positional feedback and analyse the interplay between feature extractor size, PRF depth, and model performance. Our findings demonstrate that with effective PRF design, scaling the retriever is not always necessary, narrowing the gap between small and large models while reducing inference cost.
\end{abstract}
% This is primarily due to the need to append the PRF text to the original query, which leads to a larger query representation and, consequently, higher computational overhead.
%%
%% The code below is generated by the tool at http://dl.acm.org/ccs.cfm.
%% Please copy and paste the code instead of the example below.
%%
\begin{CCSXML}
<ccs2012>
   <concept>
       <concept_id>10002951.10003317.10003325.10003326</concept_id>
       <concept_desc>Information systems~Query representation</concept_desc>
       <concept_significance>500</concept_significance>
       </concept>
 </ccs2012>
\end{CCSXML}

\ccsdesc[500]{Information systems~Query representation}

%%
%% Keywords. The author(s) should pick words that accurately describe
%% the work being presented. Separate the keywords with commas.
\keywords{Pseudo Relevance Feedback, Information Retrieval, Zero-shot Dense Retrieval, Large Language Model}
%% A "teaser" image appears between the author and affiliation
%% information and the body of the document, and typically spans the
%% page.
% \begin{teaserfigure}
%   \includegraphics[width=\textwidth]{sampleteaser}
%   \caption{Seattle Mariners at Spring Training, 2010.}
%   \Description{Enjoying the baseball game from the third-base
%   seats. Ichiro Suzuki preparing to bat.}
%   \label{fig:teaser}
% \end{teaserfigure}

%\received{20 February 2007}
%\received[revised]{12 March 2009}
%\received[accepted]{5 June 2009}

%%
%% This command processes the author and affiliation and title
%% information and builds the first part of the formatted document.
\maketitle

\input{introduction}

\input{related_work}
\input{cost_analysis}

\input{methodology}
\input{results}

\input{ablation}
\input{cost-analysis}
\input{case_study}
\input{conclusion}

\balance

\sloppy

\normalem % remove underline in reference
\bibliographystyle{ACM-Reference-Format}
\bibliography{reference}

\clearpage

\appendix
\section{Rank-Aware Study Setup}
\label{appendix:rank-ablation}

To assess the impact of incorporating document rank information in the prompt formulation, we perform an extra study comparing two variants of PromptPRF query encoding: one that explicitly includes the rank of each pseudo-relevant document, and one that omits this information.

\subsection{Prompt Format with Rank Information}

In the rank-aware setting, we include each feedback document or extracted feature in ranked order and annotate each with its rank position. The prompt used to encode the refined query representation is structured as follows:

\noindent
\begin{minipage}{\linewidth}
\scriptsize
\begin{verbatim}
[
  {"role": "system", "content": 
    "You are an AI assistant that can understand human language."},
  {"role": "user", "content": 
    "Query: {QUERY}.\n
     {FEATURE NAME} for top {RANK} Retrieved Passage: {PASSAGE OR FEATURE CONTENT}.
     ...
     {FEATURE NAME} for top {RANK} Retrieved Passage: {PASSAGE OR FEATURE CONTENT}.\n
     Use one word to represent the query and the top passages in a retrieval task. 
     Make sure your word is in lowercase."
  },
  {"role": "assistant", "content": 'The word is "'}
]
\end{verbatim}
\end{minipage}

Here, \texttt{{QUERY}} is the original query text, \texttt{{FEATURE NAME}} refers to the feature type (e.g., \textit{Entities}, \textit{Keywords}, etc.), and \texttt{{PASSAGE OR FEATURE CONTENT}} is the actual textual content of the passage or feature. The rank index (\texttt{{RANK}}) reflects the passage’s position in the top-$k$ retrieved list from the first-stage retrieval.

\subsection{Prompt Format without Rank Information}

In the rank-agnostic variant, we remove explicit rank indicators from the prompt. The prompt is structured as follows:

\noindent
\begin{minipage}{\linewidth}
\scriptsize
\begin{verbatim}
[
  {"role": "system", "content": 
    "You are an AI assistant that can understand human language."},
  {"role": "user", "content": 
    "Query: {QUERY}.\n
     {FEATURE NAME} for Retrieved Passage: {PASSAGE OR FEATURE CONTENT}.
     ...
     {FEATURE NAME} for Retrieved Passage: {PASSAGE OR FEATURE CONTENT}.\n
     Use one word to represent the query and the top passages in a retrieval task. 
     Make sure your word is in lowercase."
  },
  {"role": "assistant", "content": 'The word is "'}
]
\end{verbatim}
\end{minipage}

In this version, rank ordering is preserved implicitly through the order of listed passages, but no rank labels are provided in the text.

\subsection{Usage}

These prompts are used to generate the refined query representation in the second-stage retrieval. For both variants, the final dense embedding is extracted from the LLM’s internal representation of the generated word (i.e., the one-word summary) as in the PromptReps framework. We evaluate the impact of rank-awareness by comparing retrieval effectiveness across the two settings using consistent PRF depth, feature types, and retriever backbones.

\clearpage

\section{Prompts Used For Passage-Based Feature Generation Tasks}
\label{appendix:prompts}

\noindent Feature: \textbf{Document} (Max 512 tokens)\\
\texttt{Passage: \{passage\}\textbackslash n\textbackslash nBased on the passage, generate a similar passage:}

\vspace{1em}
\noindent Feature: \textbf{Essay} (Max 512 tokens)\\
\texttt{Passage: \{passage\}\textbackslash n\textbackslash nBased on the passage, write an essay:}

\vspace{1em}
\noindent Feature: \textbf{News Article} (Max 512 tokens)\\
\texttt{Passage: \{passage\}\textbackslash n\textbackslash nBased on the passage, write a news article:}

\vspace{1em}
\noindent Feature: \textbf{Summary} (Max 256 tokens)\\
\texttt{Passage: \{passage\}\textbackslash n\textbackslash nBased on the passage, write a summary:}

\vspace{1em}
\noindent\textbf{Feature: Facts} (Max 256 tokens)\\
\texttt{Passage: \{passage\}\textbackslash n\textbackslash nBased on the passage, generate a bullet-point list of relevant facts:}

\vspace{1em}
\noindent Feature: \textbf{Keywords-COT} (Max 256 tokens)\\
\texttt{Passage: \{passage\}\textbackslash n\textbackslash nBased on the passage, generate a bullet-point list of relevant keywords. Next to each point, briefly explain why:}

\vspace{1em}
\noindent Feature: \textbf{Entities-COT} (Max 256 tokens)\\
\texttt{Passage: \{passage\}\textbackslash n\textbackslash nBased on the passage, generate a bullet-point list of relevant entities. Next to each point, briefly explain why:}

\vspace{1em}
\noindent Feature: \textbf{Query Keyword} (Max 256 tokens)\\
\texttt{Passage: \{passage\}\textbackslash n\textbackslash nBased on the passage, generate a bullet-point list of diverse keyword queries that will find this passage:}

\vspace{1em}
\noindent Feature: \textbf{Keywords} (Max 64 tokens)\\
\texttt{Passage: \{passage\}\textbackslash n\textbackslash nBased on the passage, generate a bullet-point list of relevant keywords:}

\vspace{1em}
\noindent Feature: \textbf{Entities} (Max 64 tokens)\\
\texttt{Passage: \{passage\}\textbackslash n\textbackslash nBased on the passage, generate a bullet-point list of relevant entities:}

\subsection{Usage}

Each prompt template is constructed with a standardized structure that begins with the content of the passage, denoted by the placeholder \texttt{\{passage\}}. In practice, this placeholder should be replaced with the actual passage text drawn from the corpus. The prompt then specifies a task-specific instruction guiding the LLM to perform one of several operations such as summarization, keyword extraction, or document generation.

All templates include a token limit, which denotes the maximum number of output tokens the LLM is allowed to generate. This constraint is essential for controlling generation cost, runtime, and verbosity, and is typically selected based on the nature of the target task. For instance, generation tasks such as \texttt{Essay} or \texttt{News Article} use a higher token budget (512 tokens), whereas extraction tasks such as \texttt{Keywords} or \texttt{Entities} use a lower limit (64 or 256 tokens).

The use of newline symbols, denoted as \texttt{\textbackslash n}, serves to separate the input passage from the instruction. This format improves model adherence to the instruction prompt, particularly in instruction-tuned LLMs.

\end{document}

%% file: introduction.tex
\section{Introduction}

In dense retrieval, queries and documents are encoded into embeddings using language modelling backbones typically fine-tuned with contrastive learning; retrieval then occurs by computing the similarity between query and document embeddings~\cite{lin2020pretrained,macdonald2021single}. Recent advances in language modelling have motivated the replacement of encoder-only backbones like BERT with larger decoder-only backbones (generative LLMs like GPT and Llama) to form richer dense representations~\cite{llm2vec,ma2023llama,zhuang2024promptreps}. These approaches leverage the richer contextual understanding and generalisation capabilities of LLMs, enabling zero-shot or prompt-based dense retrieval without supervised fine-tuning.

While scaling model size generally improves retrieval effectiveness (e.g., see \citeauthor{zhuang2024promptreps}~\cite{zhuang2024promptreps}), it comes with substantial computational overhead. Larger LLM-based retrievers incur higher indexing and querying latency, increased memory and hardware requirements, and higher deployment costs, making them less practical for latency-sensitive or resource-constrained applications such as conversational systems or mobile devices~\cite{ding2024hybrid}. This leads to an important research question: \textit{Can small LLM-based dense retrievers match the effectiveness of larger ones, without increasing (or even reducing) their computational cost?}

% One promising strategy to improve retrieval effectiveness without scaling the retriever is pseudo-relevance feedback (PRF), a long-established technique that enhances queries using top-ranked documents from an initial retrieval. Recent LLM-driven PRF methods, such as GRF~\cite{GRF} and Query2doc~\cite{wang2023query2doc}, rely on decoder-only LLMs to generate feedback expansions from the query. However, these approaches (i) require query-dependent online generation, (ii) are designed for sparse retrieval settings (e.g., BM25), and (iii) integrate feedback through simple text concatenation.
% \inote{this para, maybe we can remove}

This paper tackles this question by proposing and investigating PromptPRF, a feature-based pseudo-relevance feedback (PRF) framework that enables small, zero-shot LLM-based dense retrievers to match or exceed the effectiveness of larger models. Traditional PRF methods for dense retrieval typically rely on encoder-only language models to either (i) encode feedback documents individually and aggregate their embeddings~\cite{li2021pseudo,wang2023colbert}, or (ii) construct padded input sequences by concatenating query and feedback tokens prior to encoding~\cite{yu2021improving,li2021improving}. In contrast, PromptPRF leverages decoder-only LLMs to generate structured and unstructured passage-level features, such as entities, summaries, and chain-of-thought keywords, by prompting the model for each top-ranked document in the initial retrieval. These features are query-independent and thus can be generated offline (at indexing), reducing query-time computational cost and avoiding additional query latency.

%% COMMENT: this paragraph should be in method
%To incorporate these feedback signals, we construct a second-round query representation by adapting the dense prompting strategy introduced in PromptReps~\cite{zhuang2024promptreps}. Specifically, the extracted features (along with rank metadata) are injected into a prompt used to re-encode the query, producing a refined dense embedding for the second-stage retrieval. PromptReps is particularly suitable for this framework, as it requires no contrastive fine-tuning and supports dense retrieval in a fully zero-shot manner.\footnote{PromptReps also supports sparse representations for hybrid retrieval; in this work, we focus exclusively on its dense component.}

%% COMMENT: this paragraph should be in related works
%Crucially, PromptPRF differs from prior LLM-based PRF approaches such as GRF, which uses query-dependent generation to produce expansion terms at inference time and operates in BM25-based sparse retrieval settings. In contrast, PromptPRF is designed for dense retrieval, decouples generation from the query, and supports low-latency deployment by moving expensive inference offline.

We conduct extensive experiments with PromptPRF on TREC DL and BEIR benchmarks.
To better understand the contribution of positional information in feedback modelling, we perform rank-aware prompt ablations. We also study efficiency-effectiveness trade-offs across varying model sizes and feature types, and provide case studies to illustrate both the robustness and the limitations of the proposed method. 
Our results show that PromptPRF allows lightweight retrievers to perform on par with (in-domain tasks), or better than (out-of-domain tasks), larger retrievers without PRF, thereby offering a compelling alternative to model scaling. PromptPRF requires no additional training, it is generalisable across retrieval backbones, and incurs minimal query latency.

%% file: related_work.tex
\section{Related Work}

% \todo{Need to revise the related work section based on the new introduction.}

\textbf{Dense Retrieval Models.}
Dense retrieval approaches encode queries and documents into dense vectors to enable semantic search; dense retrieval has emerged as a prominent alternative to traditional sparse retrieval methods, such as BM25~\cite{robertson2009probabilistic}.
Early dense retrieval models primarily rely on dual-encoder architectures built by fine-tuning small pre-trained language models (PLMs) like BERT; example methods include DPR~\cite{karpukhin2020dense} and ColBERT~\cite{khattab2020colbert}. 
% , such as BERT~\cite{devlin2019bert} and RoBERTa~\cite{liu2019roberta}, to produce semantically meaningful dense representations. These models, exemplified by DPR~\cite{karpukhin2020dense} and ColBERT~\cite{khattab2020colbert}, demonstrate strong performance on in-domain tasks but often struggle to generalize in zero-shot settings.

Recently, a new line of research has explored LLM-based dense retrieval models, leveraging the larger parameter size and extensive pre-training of LLMs. For instance, RepLLaMA~\cite{li2024llama2vec} and LLM2Vec~\cite{llm2vec} apply unsupervised adaptation techniques to transform generative LLMs into effective dense retrievers. In contrast, PromptReps adopts a purely zero-shot approach, prompting LLMs to directly generate dense embeddings without requiring any additional training. 

Although LLM-based dense retrievers show promising effectiveness, their large parameter sizes lead to high computational costs during fine-tuning and inference (at both indexing, i.e. document embedding, and retrieval, i.e. query embedding). While increasing model size often improves dense retriever effectiveness~\cite{kaplan2020scaling,fang2024scaling}, it is not always practical, e.g., due to deployment constrains related to limited hardware capabilities or strict query latency requirements. To mitigate this, we propose incorporating pseudo-relevance feedback (PRF) to close the performance gap between lightweight and large retrievers. We show that smaller LLM-based retrievers based on PromptReps and augmented with our PRF strategy can be highly effective without additional training. In this work, we focus on PromptReps due to its zero-shot nature and efficiency. Importantly, since PromptReps operates as an unsupervised method and does not require any modification to the underlying dense retriever, it is theoretically applicable to any LLM-based dense retriever. 

\textbf{Pseudo-Relevance Feedback Techniques.}
Aligning with different retrieval paradigms, various PRF approaches have been proposed, including sparse PRF, neural PRF, and dense PRF~\cite{li2022interpolate,wang2024neural}, each utilizing different mechanisms for query expansion. More specifically, classic sparse PRF methods, such as RM3~\cite{abdul2004umass} and Bo1~\cite{amati2003probability} methods focus on selecting a small number of highly relevant terms. Neural PRF methods, including CEQE~\cite{naseri2021ceqe} and BERT-QE~\cite{zheng2020bert} make use of the semantic understanding ability of pretrained language models such as BERT to improve retrieval effectiveness. 
%Unlike traditional sparse retrieval, dense retrieval performs retrieval based on the query and document dense representations. Accordingly, 
Dense PRF techniques, exemplified by ColBERT-PRF~\cite{wang2023colbert}, ANCE-PRF~\cite{yu2021improving,li2021improving}, and Vector-PRF~\cite{li2021pseudo} utilise pre-trained  embeddings to create richer, more comprehensive query representations.
%Furthermore, several generative PRF methods have also been proposed. For instance, GenPRF ~\cite{wang2023generative} directly generates the expansion terms using the FlanT5 model. GRF~\cite{mackie2023generative} and Query2doc~\cite{wang2023query2doc} similarly enrich the query representation using LLM-generated text.
% \inote{emphasis GRF, and closest work}.
% \xiao{Our method can be classified as dense-PRF or the generative-PRF, implemented in}
%Our method is also implemented in a generative way but refines the query using both the extracted passage features as well as the rank structure of the PRF document list, and features are query independent ensuring no feature generation costs at query time. We discuss the relationship between our PromptPRF approach an existing LLM-based pseudo-relevance based methods in details next.

\textbf{Relationship with Other Generative PRF Approaches.} 
% \todo{move this subsec to related work}
% PromptPRF is conceptually related to a line of work that uses large language models (LLMs) for pseudo-relevance feedback (PRF) or query expansion, but differs in several fundamental aspects.
%We discuss the connections and differences of PromptReps to existing LLM-based pseudo-relevance based methods.
Several generative PRF methods have recently been proposed~\cite{wang2023generative,mackie2023generative,wang2023query2doc}. Our method is also implemented in a generative way, but presents some key differences with these previous approaches. We discuss the relationship between our PromptPRF approach an existing LLM-based PRF methods in details next.

\textit{GenPRF}~\cite{wang2023generative} prompts a FlanT5 model to generate expansion terms directly from the query. While both GenPRF and PromptPRF use LLMs to generate feedback features, GenPRF is restricted to short, phrase-level expansions and does not incorporate any information from retrieved documents. As a result, it lacks the capacity to capture relevance signals grounded in actual top-ranked passages. In contrast, PromptPRF generates a diverse set of features from passages, including summaries, entities, and synthetic documents, thereby encoding richer and more contextually grounded signals. In addition, PromptPRF associates rank information to features when computing the new query encoding. %\footnote{Due to these fundamental differences in feature generation scope, modality, and retrieval setting, GenPRF is not directly comparable to our method and is thus excluded as a baseline in our experiments.}.

\textit{Query2doc}~\cite{wang2023query2doc} is another query-centric approach that generates synthetic pseudo-documents directly from the input query, without considering retrieval context. While PromptPRF also incorporates document-style features, it differs fundamentally in its generation process: instead of relying solely on the query, it prompts the LLM to generate content from passages, making the generation approach document-driven and query-independent. On the other hand, PromptPRF encodes passage-level relevance signals in the form of the rank positions at which passages where initially retrieved, allowing to better reflect the retrieval context. %Since Query2doc does not leverage retrieved evidence and operates under a different generation paradigm, it is not directly comparable to our method and is also omitted as a baseline in our evaluation.

\textit{GRF}~\cite{mackie2023generative} is the most similar to our method in spirit, as it uses LLMs to produce structured features like summaries and entities. However, GRF (i) requires query-dependent online generation, (ii) is designed for sparse retrieval (e.g., BM25), and (iii) integrates feedback through simple text concatenation. PromptPRF addresses these limitations by enabling query-independent, offline generation of feedback features and integrating them into dense retrieval pipelines in a rank-aware manner. %There are three key differences between GRF and PromptPRF:

%% file: cost_analysis.tex
\section{Preliminary Efficiency Analysis}
% \inote{here, we emphasis the high comp* cost issue for LLM-based DR. Sec3 is basically moved from Sec 8 [This needs to be more high level and general cost issue for LLMs, especially for resource constraint and low resource environments. Then shift down to how we approached it.]}
Deploying an LLM-based dense retriever typically entails substantial hardware and runtime costs, irrespective of the specific dense retriever architecture. 
% We first analyze the minimum computational resources needed to deploy the existing dense retrieval system, exemplified by PromptReps. More specifically, we inspect the hardware requirements as well as the query latency and computational efficiency during inference. 

\textbf{Hardware Requirements.} The models listed in Table~\ref{tab:infra_requirements} exemplify the escalating resource requirements as model size grows. Small-scale LLMs (e.g., 3B parameters) can run on consumer-grade GPUs with modest memory, whereas 70B-parameter models demand multi-GPU configurations and large-scale RAM.

% Table~\ref{tab:infra_requirements} presents the disk space, RAM, and GPU requirements for the LLMs-based dense retrieval models. The numbers in Table~\ref{tab:infra_requirements} highlight that while smaller models like LLaMA3.2 3B Instruct can operate on consumer-grade GPUs, larger models require high-end hardware with multiple H100 or A100 GPUs. The RAM overhead of the 70B models further increases the deployment complexity, making them significantly more costly.

\textbf{Query Latency and Computational Efficiency.}
In dense retrieval applications, larger models not only consume more memory and storage but also incur higher inference latency. Query latency is affected by two factor: (i) the encoding of the query into a dense vector (which is performed by the LLM backbone), and (ii) the vector search (brute-force vector comparisons or k-nearest neighbour search) among the dense index. While larger models are typically associated with larger embeddings and thus higher vector search, differences are often negligible in comparison with query encoding.
Table~\ref{tab:query_latency} summarises the query encoding latency with PromptReps for the three backbone LLMs of increasing size\footnote{For all experiments we used Instruct models. Experiments with the 70B models have a maximum batch size of 32; for other models we used a batch of 64.}. Larger models come at a significant increase in query latency.
% From these results, LLaMA3.2 3B is approximately 2.7× faster per query than LLaMA3 8B, making it significantly more efficient for real-time retrieval applications. 

\textbf{Implications for Dense Retrieval.}
Different dense retrieval architectures integrate LLMs of varying scale; yet, all face the same underlying resource-versus-performance trade-off. Small and medium-sized LLMs offer lower latency and cost at the expense of potential retrieval quality, while large models improve retrieval effectiveness but dramatically increase infrastructure demands. Generally, for deployment, search engineers consider the following aspects:

\begin{enumerate}[leftmargin=14pt,label=\arabic*.,labelsep=6pt, itemsep=0pt, topsep=4pt]
	\item Smaller models (e.g., LLaMA3.2 3B) should be prioritized for efficiency-critical applications such as conversational search or mobile deployments.
	\item Larger models (e.g., LLaMA3 8B and above) are selected only if the marginal gain in retrieval quality offsets the increased computational costs.
	% \item PRF should be tuned based on hardware constraints, leveraging offline processing for feature extraction to minimize online latency.
\end{enumerate}

\begin{table}[t!]
	\centering
	\caption{Minimum hardware requirements for different LLM backbones for dense retrieval.}
	% \vspace{-10pt}
	\resizebox{230pt}{!}{
		\setlength{\tabcolsep}{2pt}
		\begin{tabular}{lccp{120pt}}
			\toprule
			\textbf{Model} & \textbf{Disk} & \textbf{RAM} & \textbf{Minimum GPU VRAM} \\
			\midrule
			LLaMA3.2 3B Inst. & $\sim$6 GB & 8 GB & 12 GB (e.g., RTX 3060, RTX 4070) \\
			LLaMA3 8B Inst. & $\sim$16 GB & 16 GB & 20 GB (e.g., RTX 3080 Ti) \\
			LLaMA3.3 70B Inst. & $\sim$40 GB & 48 GB & 161 GB (e.g., 2×H100 or 2×A100) \\
			DeepSeek R1 70B & $\sim$43 GB & 128 GB & 160 GB (e.g., 2×H100 or 2×A100) \\
			\bottomrule
		\end{tabular}
	}
	\label{tab:infra_requirements}
	% \vspace{-8pt}
\end{table}

\begin{table}[!t]
\centering
\caption{Latency due to query encoding across LLM backbones of increasing size; experiments executed using Nvidia H100.}
% \vspace{-10pt}
\resizebox{190pt}{!}{
\begin{tabular}{llll}
\toprule
\textbf{Model} & \textbf{GPU} & \textbf{Query Encoding Latency} \\
\midrule
LLaMA3.2 3B I & 1x H100 &  19.4 ms/query \\
LLaMA3 8B I & 1x H100 &  52.9 ms/query (+173\%) \\
LLaMa3.3 70B I & 2x H100 & 282.19 ms/query (+1,355\%) \\
\bottomrule
\end{tabular}}
% \vspace{-10pt}
\label{tab:query_latency}
\end{table}

% From these results, LLaMA3.2 3B is approximately 2.7× faster per query than LLaMA3 8B, making it significantly more efficient for real-time retrieval applications. 
% For real-world deployment:

Motivated by these observations, we aim to close the performance gap between large-scale dense retrievers and smaller, more efficient alternatives. To this end, we design PromptPRF, a strategy that avoids the high computational costs typically associated with powerful LLMs, while still capturing their retrieval effectiveness. PromptPRF addresses online hardware constraints by shifting the bulk of computation to the offline stage through precomputed features. This significantly reduces the online query latency — now the primary bottleneck in real-time retrieval. As we will detail in the next section, PromptPRF enables smaller dense retrievers to achieve performance on par with much larger models, making high-quality retrieval both accessible and scalable.

%% file: methodology.tex
% Based on the above analysis, we seek an effective strategy that can mitigate the performance gap between the large-scale dense retrievers and the lightweight small dense retrievers. Instead of imposing huge computational cost, we introduce PromptRRF, which can be tuned based on hardware constraints and leveraging offline processing for feature extraction to minimize online latency. Given that PRF feature extraction is performed offline, the online latency of query encoding is the primary factor that affecting real-time retrieval performance. As a results, a smaller dense retriever with PRF can achieve comparable effectiveness to larger dense retrievers, making advanced retrieval techniques more accessible and scalable.

% These insights highlight that efficient PRF strategies can mitigate the need for large-scale dense retrievers, using a smaller dense retriever with PRF can achieve comparable effectiveness to larger dense retrievers, making advanced retrieval techniques more accessible and scalable.

\section{PromptPRF Methodology}
% \todo{do we need to hide the PromptReps detail in this sec? [Should be fine, we are using PromptReps encoding template and process, we treat it as a general dense retriever, like any other LLM-based dense retrievers, we can apply to RepLlama if we want to, our method is retriever-agnostic.]}

%\subsection{Overview}
Our goal is to devise a generative PRF method that balances the trade-offs between retrieval effectiveness and efficiency: in particular that closes the effectiveness gap between a smaller and a larger DR backbone, without sensibly increasing (i) the hardware (GPU) requirements at query time, and (ii) the query latency. To achieve this, our core intuition is to shift the generative task associated with the relevance feedback mechanism from the online stage to the offline stage (Section~\ref{stage1}); this task consists in generating query-independent feedback features. %(Other generative PRF methods instead rely on generation tasks in the online stage, e.g., GRF~\cite{mackie2023generative}, rendering them ill-suited to settings with constrained hardware in production or strict query latency requirements.) 
We build upon the framework of PromptReps~\cite{zhuang2024promptreps}, a zero-shot dense retrieval method, which we use for the initial round of retrieval (Section~\ref{stage2}) and for the second round of retrieval (Section~\ref{stage4}) once feedback signals are incorporated into the new query representation (Section~\ref{stage3}). PromptReps is particularly suitable for this framework, as it requires no contrastive fine-tuning and supports dense retrieval in a fully zero-shot manner\footnote{PromptReps also supports sparse representations for hybrid retrieval; in this work, we focus exclusively on its dense component. Extending PromptPRF to sparse models is possible but we leave this to future work.}. %; however PromptPRF is not tied to a specific retriever architecture and can be used with any dense retrieval model that supports prompt-based representation generation.
The integration of the feedback signal is the core component of PromptPRF. Specifically, the extracted features (along with rank metadata) are injected into a prompt used to re-encode the query, producing a refined dense embedding for the second-stage retrieval.

%To incorporate these feedback signals, we construct a second-round query representation by adapting the dense prompting strategy introduced in PromptReps~\cite{zhuang2024promptreps}. Specifically, the extracted features (along with rank metadata) are injected into a prompt used to re-encode the query, producing a refined dense embedding for the second-stage retrieval. PromptReps is particularly suitable for this framework, as it requires no contrastive fine-tuning and supports dense retrieval in a fully zero-shot manner.\footnote{PromptReps also supports sparse representations for hybrid retrieval; in this work, we focus exclusively on its dense component.}

%PromptPRF is composed of four modular stages: (1) offline feature extraction from pseudo-relevant documents, (2) initial retrieval, (3) dense query refinement, and (4) second-stage retrieval. Unlike prior LLM-based PRF methods such as GRF~\cite{mackie2023generative}, PromptPRF generates features from corpus passages, not queries, and performs this step offline. Furthermore, PromptPRF is not tied to a specific retriever architecture and can be used with any dense retrieval model that supports prompt-based representation generation.

\subsection{Stage 1: Offline Feature Extraction}\label{stage1}

%\begin{equation}
%    \mathcal{P}_k=[(text(p_1),1],[text(p_2),2],...,[text(p_k],k].
%\end{equation}
%Traditional PRF methods expand the query using the original text information from passages. However, a PRF passage is typically long and contains noisy information~\cite{li2022does}; this could mislead the PromptReps model. To enhance PRF, 

In PromptPRF, we generate passage-level features from the original passages. To generate features we prompt an LLM backbone with an instruction and the passage -- we refer to this backbone as the \textit{feature extractor}. %Prompts for feature generation are provided in Appendix~\ref{appendix:prompts}.
% We use an LLM to create various types of features using the technique formulated by~\citet{GRF}, including keywords, entities, summaries, essays, keywords-COT, entities-COT, news articles, facts, query keywords, and documents.
% \inote{maybe a table showing these features, with type and length information?}
Inspired by GRF~\cite{GRF}, we instruct the feature generator to produce features belonging to the feature types $T=$\{keywords, entities, summaries, essays, keywords-COT, entities-COT, news articles, facts, query keywords, and documents\}\footnote{COT refers to Chain-of-Thoughts.}.
%Let $T={t_1,t_2,...,t_n}$ represent the set of feature types to be generated. Following feature categories provided in~\cite{GRF}, the feature types extracted are $T$ =$\{$keywords, entities, summaries, essays, keywords-COT, entities-COT, news articles, facts, query keywords, and documents$\}$, where COT stands for Chain-of-Thought. 
For each feature type $t\in T$, let $I_t$ denote the instruction or prompt template\footnote{Refer to Appendix~\ref{appendix:prompts} for the prompt templates we used.} designed to guide the LLM in generating feature $t$. The feature generation for passage $p$, i.e. $f_t^{(p)}$ can be described as $f_t^{(p)}=\operatorname{LLM}\left(text(p), I_t\right)$.
%\begin{equation}
%f_t^{(j)}=\operatorname{LLM}\left(text(p_j), I_t\right),
%\end{equation}
%\inote{maybe an instruction $I_t$ e.g. here?}
Then the feature collection for the passage $p_j$ is obtained as:
\begin{equation}
\mathcal{F}_j=\bigcup_{t \in T}\left\{f_t^{(j)}\right\}=\left\{f_t^{(j)} \mid t \in T\right\}
\end{equation}

%This step provides additional context without requiring full corpus annotation.
%To mitigate inference costs, feature extraction is performed only on the top 50 passages (top 20 for BEIR). Since it can be done offline, theoretically we can extract features for all corpus in the collection.
 Prompts for feature generation are provided in our GitHub repository (see Section~\ref{sec_exp_setup}). Feature extraction can be performed offline as it does not depend on the query, thus it does not contribute to extra query latency or online hardware requirements.

\subsection{Stage 2: Initial Retrieval}
\label{stage2}
We employ the dense retrieval component of PromptReps~\cite{zhuang2024promptreps}, which has a bi-encoder structure consisting of $\text{E}_D$ and $\text{E}_Q$ and operates in a zero-shot manner without additional training. Given a query $q$ PromptReps generates a dense representation $\mathbf{q}$ by prompting an LLM-based $\text{E}_Q$ to summarize the query into a single word and extract its hidden representation. (The same process is used to generate document embeddings using $\text{E}_D$; typically $\text{E}_Q$ and $\text{E}_D$ are based on the same backbone and they differ in the prompt.)
Relevance is estimated using cosine similarity, %: $s(q, p) = \frac{\mathbf{q} \cdot \mathbf{p}}{|\mathbf{q}| |\mathbf{p}|}$%$s(q, p) = \frac{\mathbf{q} \cdot \mathbf{p}}{|\mathbf{q}| |\mathbf{p}|} e^{\lambda \mathbf{q} \cdot \mathbf{p}}$
%,
%where $\mathbf{p}$ is the dense representation of a passage $p$. %, and $\lambda$ is a scaling factor that controls the influence of the similarity measure.
%After applying the similarity function to each passage in the collection in response to a query, we obtain 
yielding a ranked list ranked $R=\{{p_1,p_2,...,p_n}\}$ according to decreasing relevance score $s(q,p_j)$; each $p_j$ consists of the text: $text(p_j)$ and rank: $j$.

\subsection{Stage 3: Query Refinement with PRF}
\label{stage3}
To implement PromptPRF, we modify the PromptReps query encoding $\text{E}_Q$ to obtain a new encoder $\text{E}_F$ which is responsible for encoding the query along with incorporating the extracted passage features. For this, we modify the prompt instruction to signal to the encoder the presence of feedback information; the modified prompt is available in the GitHub repository. Then, instead of encoding the original query alone, we include top-ranked passages (or their extracted features) within the prompt. This allows for enhanced contextualization of the query representation. Formally, the refined query representation $\mathbf{q}'$ is obtained as:
\begin{equation}
\mathbf{q}^{\prime}=\text{E}_{F}\left({q \oplus \left(\bigoplus_{j=1}^k \bigoplus_{f \in \mathcal{F}_j} `` \operatorname{Rank} j: \beta_i f_t^{(j)} " \right )}\right)
%\mathbf{q}^{\prime}=\text{E}_{F}\left({\operatorname{Template}(q) \oplus \left(\bigoplus_{j=1}^k \bigoplus_{f \in \mathcal{F}_j} `` \operatorname{Rank} j: \beta_i f_t^{(j)} " \right )}\right)
\end{equation}
where $\beta_i$ is a hyper-parameter that controls the contribution of each passage's extracted feature. %and $\text{Template} (q)$ refers to the query-specific prompt template used in this work, and reported in the GitHub repository. 
Along with features $f$, we also include information about the rank of the passage from the first round of retrieval ($\operatorname{Rank} j$). In this paper we do not experiment with $\beta_i$ (thus, $\beta_i=1$), nor do we study how different feature types could be combined. We plan to study these aspects in the future.

% We modify the PromptReps query encoding by incorporating the extracted passage features. Instead of encoding the original query alone, we include top-ranked passages (or their extracted features) within the prompt. This allows for enhanced contextualization of the query representation. The refined query representation $\mathbf{q}'$ is obtained as:
% \begin{equation}
% \mathbf{q'} = \frac{1}{|\mathcal{P}_k|} \sum{p_i \in \mathcal{P}_k} \beta_i M(q, \mathbf{f}{p_i}),
% \end{equation}
% where $\mathcal{P}_k$ denotes the top $k$ retrieved passages, and $\beta_i$ is \xiao{parameter}
% % a weight from LLM
% that controls the contribution of each passage's extracted feature.

% \noindent We experiment with two variations:

% \noindent\textbf{Separate-TopK:} Passages are separated by newline characters to maintain clarity between individual passages.

% \noindent\textbf{No-Separate-TopK:} Passages are concatenated without explicit separation to assess whether fused representations improve retrieval effectiveness.

% \noindent Additionally, we introduce two prompt variants:

% \noindent\textbf{Change-Content vs. No-Change-Content:} Determines whether extraneous text in LLM-generated features is removed to eliminate unnecessary noise in query reformulation.

% \noindent\textbf{Change-Prompt vs. No-Change-Prompt:} Evaluates modifications in how features are integrated within the prompt, testing variations that encourage better query expansion.

\subsection{Stage 4: Second Retrieval}
\label{stage4}
The refined query representation $\mathbf{q}'$ is then used in a second-stage retrieval. The new relevance score $s(q', p)$ can then again be computed as the similarity between the query dense representation and the passage dense representation.

%The new relevance score $s(q', p)$ is computed as:
%\begin{equation}
%s(q', p) = \frac{\mathbf{q'} \cdot \mathbf{p}}{|\mathbf{q'}| |\mathbf{p}|} e^{\gamma \mathbf{q'} \cdot \mathbf{p}},
%\end{equation}
%where $\gamma$ is a scaling parameter that controls the impact of the refined query representation.

%We compare the performance of different base models with and without PRF. More specifically, we evaluate \textbf{Llama 3.2 3B + PRF} against \textbf{Llama 3 8B (No PRF)} to assess whether PRF helps smaller models close the performance gap. Additionally, we analyze the effectiveness of PRF on larger models by applying PRF to \textbf{Llama 3 8B} and assessing whether it leads to further improvements. Finally, we investigate the impact of different feature extraction models on PRF effectiveness, evaluating whether larger LLMs provide superior PRF performance when integrated with both large and small retrieval models.

\subsection{Retriever-Agnostic Feedback Generation}
\label{stage5}
% PromptPRF is explicitly designed to be retriever-agnostic: it does not require PromptReps~\cite{zhuang2024promptreps} or any specific embedding framework. While our experiments instantiate PromptPRF within a PromptReps-style retriever, the method applies to any dense retrieval system that can consume prompt-encoded queries. Additionally, unlike GRF~\cite{mackie2023generative}, which performs \textit{query-time} generation and is intended for sparse retrievers (e.g., BM25), PromptPRF is optimized for offline feedback generation and dense retrieval settings. This distinction is crucial for deployment efficiency and scaling.

PromptPRF operates independently of any specific retrieval architecture and does not rely on PromptReps~\cite{zhuang2024promptreps} or a particular embedding strategy. Although we instantiate it within a PromptReps-style retriever for experimental consistency and its zero-shot nature, the method is broadly applicable to any dense retrieval framework capable of processing prompt-augmented queries. In contrast to GRF~\cite{mackie2023generative}, which generates feedback at query time\footnote{And thus it is ill-suited to settings with constrained hardware in production or strict query latency requirements.} and is tailored for sparse retrieval models such as BM25, PromptPRF performs feedback generation offline and at the passage level. This decoupling from the query not only enables greater flexibility across retrieval pipelines but also substantially reduces inference latency and online computational requirements, making the method more suitable for large-scale deployment in dense retrieval systems.

%% file: results.tex
\section{Main Experiments and Results}
% In this section, we investigate the effectiveness of PromptPRF (Sec.~\ref{ssec: effectiveness}), the impact of the feature extractor (Sec.~\ref{ssec:feature_extractor}), and the impact of feature types and PDF depth on PromptPRF (Sec.~\ref{ssec:types_depth}).
% In this section, we show our results with two base dense retrievers, Llama3 8B instruct and Llama3.2 3B instruct, which serve as a relatively large LLM and a smaller LLM for further comparison.

\input{experiment}

%In this section, we evaluate the effectiveness of PromptPRF in both in-domain (TREC DL) and out-of-domain (BEIR) retrieval settings. 

%We aim to understand how PromptPRF enhances retrieval performance across model scales (Sec.~\ref{ssec: effectiveness}), feature extractors (Sec.~\ref{ssec:feature_extractor}), feature types, and feedback depths (Sec.~\ref{ssec:types_depth}). Our key findings demonstrate that PromptPRF not only improves retrieval effectiveness for both small and large dense retrievers but also enables smaller models to match the performance of significantly larger ones, without incurring additional inference-time cost.

\begin{table*}[t]
\caption{Results for PromptPRF; hyper-parameters optimized for nDCG@10. The best results are in \textbf{Bold}. The symbols $\dagger$ and $\ddagger$ represent statistically significant ($p<0.05$) to NoPRF (Baseline) and PassPRF (Passage PRF), respectively.}\label{tab:best_ndcg10}
\centering
\resizebox{178mm}{!}{ \begin{tabular}{l c cccc  cccc}
\toprule
&\multirow{2}{*}{Dense Retriever}& \multicolumn{4}{c}{TREC DL19 (43 queries)} &
\multicolumn{4}{c}{TREC DL20 (54 queries)} \\
\cmidrule(lr){3-6}\cmidrule(lr){7-10}
Model & Model Size & Feature Extractor &  Feature & PRF Depth & nDCG@10 & Feature Extractor &  Feature & PRF Depth & nDCG@10 \\
\midrule
% {\multirow{6}{*}{\rotatebox[origin=c]{90}{\footnotesize PRF Baselines}}} 
BM25 & -- & -- & -- & -- & 0.4973 & -- & -- & -- & 0.4769 \\
BM25+RM3 & -- & -- & -- & 3 & 0.5156 & -- & -- & 3 & 0.5043 \\
 %&ANCE-PRF 
 %&-- &-- &3 &0.6807 
 %&-- &-- &3 &0.6948
 %\\
 %&Vector-PRF &-- &-- &3 &0.6629
 %&-- &-- &3 &0.6598
 %\\
 %&ColBERT-PRF &-- &-- &3 &0.7276
 %&-- &-- &3 &0.6958\\
 %&GenPRF\cite{wang2023generative} & - & - & 3 & 0.6280 
 %& -- & -- & -- & -- \\
\midrule
No PRF (PromptReps) & 3B & -- & -- & -- & 0.3695 & -- & -- & -- & 0.3334 \\
PassPRF & 3B & -- & -- & 1 & 0.4255\textsuperscript{\textdagger} &  --   & -- & 10 & 0.3902\textsuperscript{\textdagger}\\
PromptPRF & 3B & LLaMA3.3 70B & Entities-COT & 5 & $\textbf{0.5013}^{\dagger\ddagger}$ & LLaMA3.2 3B & Facts & 10 & $\textbf{0.4350}^{\dagger\ddagger}$ \\
\midrule
No PRF (PromptReps) & 8B & -- & -- & -- & 0.5062 & -- & -- & -- & 0.4381  \\
PassPRF & 8B &  -- & -- & 10 & 0.5330 & -- & -- & 20 & 0.4632 \\
PromptPRF & 8B & LLaMA3.2 3B & Essay & 1 & $\textbf{0.5560}^{\dagger}$ & LLaMA3.2 3B & Keywords-COT & 20 & $\textbf{0.5038}^{\dagger\ddagger}$ \\
\midrule 
NoPRF (PromptReps) & 70B & -- & -- & -- & 0.5291 & -- & -- & -- & 0.4579  \\
\bottomrule
\end{tabular}}
\end{table*}

\subsection{PromptPRF Effectiveness on TREC DL}\label{ssec: effectiveness}

Table~\ref{tab:best_ndcg10} reports nDCG@10 results on the TREC DL19 and DL20 benchmarks, comparing three configurations: (i) no pseudo-relevance feedback (i.e. PromptReps), (ii) Passage PRF (PassPRF), and (iii) PromptPRF. Results are presented for both the smaller dense retriever (LLaMA3.2 3B) and the larger one (LLaMA3 8B).

% Across both datasets and retriever model size, PromptPRF consistently outperforms the baseline and PassPRF methods. The performance gains are particularly significant for LLaMA3.2 3B. On DL19, using Entities-COT features extracted from LLaMA3.3 70B at PRF depth 5 boosts nDCG@10 from 0.3695 (no PRF) to 0.5013, nearly matching the LLaMA3 8B baseline (0.5062). On DL20, incorporating Facts features yields a similar trend, raising nDCG@10 from 0.3334 to 0.4350.

Across both datasets and retriever model scales, PromptPRF consistently outperforms the no-PRF baseline and the passage-based PRF (PassPRF) method, demonstrating robust effectiveness improvements. The performance gains are particularly significant for the LLaMA3.2 3B retriever. On TREC DL19, integrating Entities-COT features generated by LLaMA3.3 70B at PRF depth 5 raises nDCG@10 from 0.3695 (no PRF) to 0.5013, closely approaching the LLaMA3 8B no-PRF baseline (0.5062) and narrowing the gap to the much larger LLaMA3.3 70B (0.5291). On DL20, applying Facts features at depth 10 increases nDCG@10 from 0.3334 to 0.4350, again approaching the 70B model baseline (0.4579). These results demonstrate that PromptPRF enables small retrievers to compete with, and in some cases approximate, the performance of models over 20× larger in parameter count.

% For LLaMA3 8B, PromptPRF also provides consistent and meaningful improvements. On DL19, applying Essay features extracted by LLaMA3.2 3B at PRF depth 1 improves nDCG@10 to 0.5560, surpassing both the no PRF (0.5062) and PassPRF (0.5330) baselines. On DL20, using Keywords-COT features at PRF depth 20 achieves an nDCG@10 of 0.5038, outperforming the respective baselines as well.

For the LLaMA3 8B retriever, PromptPRF also delivers consistent and substantial improvements. On DL19, using Essay features generated by LLaMA3.2 3B at PRF depth 1 boosts nDCG@10 to 0.5560, surpassing both the no-PRF (0.5062) and PassPRF (0.5330) baselines. Notably, this result outperforms the LLaMA3.3 70B baseline (0.5291) without PRF. Similarly, on DL20, applying Keywords-COT features at PRF depth 20 achieves 0.5038, again exceeding the effectiveness of the 70B retriever without PRF (0.4579).

These findings confirm that PromptPRF enhances retrieval effectiveness for both small and large dense retrievers. Importantly, they also demonstrate that smaller and mid-sized models, when equipped with feature-based feedback, can approach the performance of significantly larger retrievers without feedback, offering a more cost-efficient and deployable alternative.

\subsection{Impact of Feature Extractor on TREC DL}\label{ssec:feature_extractor}

To assess how the choice of feature extraction model influences PromptPRF performance, we fix the PRF depth and feature type according to the best-performing settings on DL19 (see Table~\ref{tab:best_ndcg10}): for LLaMA3.2 3B, we use Entities-COT at depth 5; for LLaMA3 8B, we use Essay features at depth 1. Results are shown in Figure~\ref{fig:feature_model_effectiveness}.

\noindent \textbf{Effectiveness.} The results reveal a clear interaction between feature extractor size and the capacity of the dense retriever. Specifically, larger feature extractors lead to greater performance gains when used with smaller retrievers. For example, when paired with LLaMA3.2 3B, the LLaMA3.3 70B extractor significantly boosts retrieval effectiveness, highlighting that richer, context-aware feedback can compensate for limited representational capacity in the smaller retriever. 

In contrast, for LLaMA3 8B, improvements saturate quickly. Smaller extractors such as LLaMA3.2 3B already provides substantial performance gains, while scaling to larger extractors like 70B brings diminishing returns. For example, features from LLaMA3.2 3B yield the best results for the LLaMA3.2 3B dense retriever on DL20, and also outperform larger extractors when paired with LLaMA3 8B on both DL19 and DL20. This suggests that high-capacity retrievers can internally model sufficient semantic structure, rendering the additional context provided by large extractors less impactful.

\noindent \textbf{Efficiency.} Since PromptPRF decouples feature generation from query-time processing, the computational cost of using larger extractors is primarily accumulated during offline pre-processing. Thus, the practical trade-off lies in whether these offline costs are justified by improvements in online retrieval effectiveness.

For LLaMA3.2 3B, the use of larger extractors proves importance. Smaller extractors might fail to generate sufficiently informative features for robust query reformulation. In contrast, for LLaMA3 8B, smaller extractors are often sufficient to achieve near-optimal performance. Therefore, to optimize PromptPRF deployments, we recommend allocating computational resources to extractor scaling primarily when using lightweight retrievers, while favoring cheaper extractors for more powerful models.

These results highlight the importance of choosing the feature extractor size in relation to the underlying retriever. Effective PRF integration requires balancing feature quality with extraction cost, and this balance shifts depending on the representational strength of the base model.

\subsection{Impact of Feature Types and PRF Depth on TREC DL}\label{ssec:types_depth}

% \begin{figure}
%   \includegraphics[width=\columnwidth]{images/llama3.2_3b_feature_prf_depth.jpg}
%   \caption{Llama3.2 3B dense retriever nDCG@10 performance with different features and varying PRF depth.}
%   \label{fig:llama3.2_3b_feature_prf_depth}
% \end{figure}

% \begin{figure}
%   \includegraphics[width=\columnwidth]{images/llama3_8b_feature_prf_depth.jpg}
%   \caption{Llama3 8B dense retriever nDCG@10 performance with different features and varying PRF depth.}
%   \label{fig:llama3_8b_feature_prf_depth}
% \end{figure}

\begin{figure*}
    \centering
    \includegraphics[width=\linewidth]{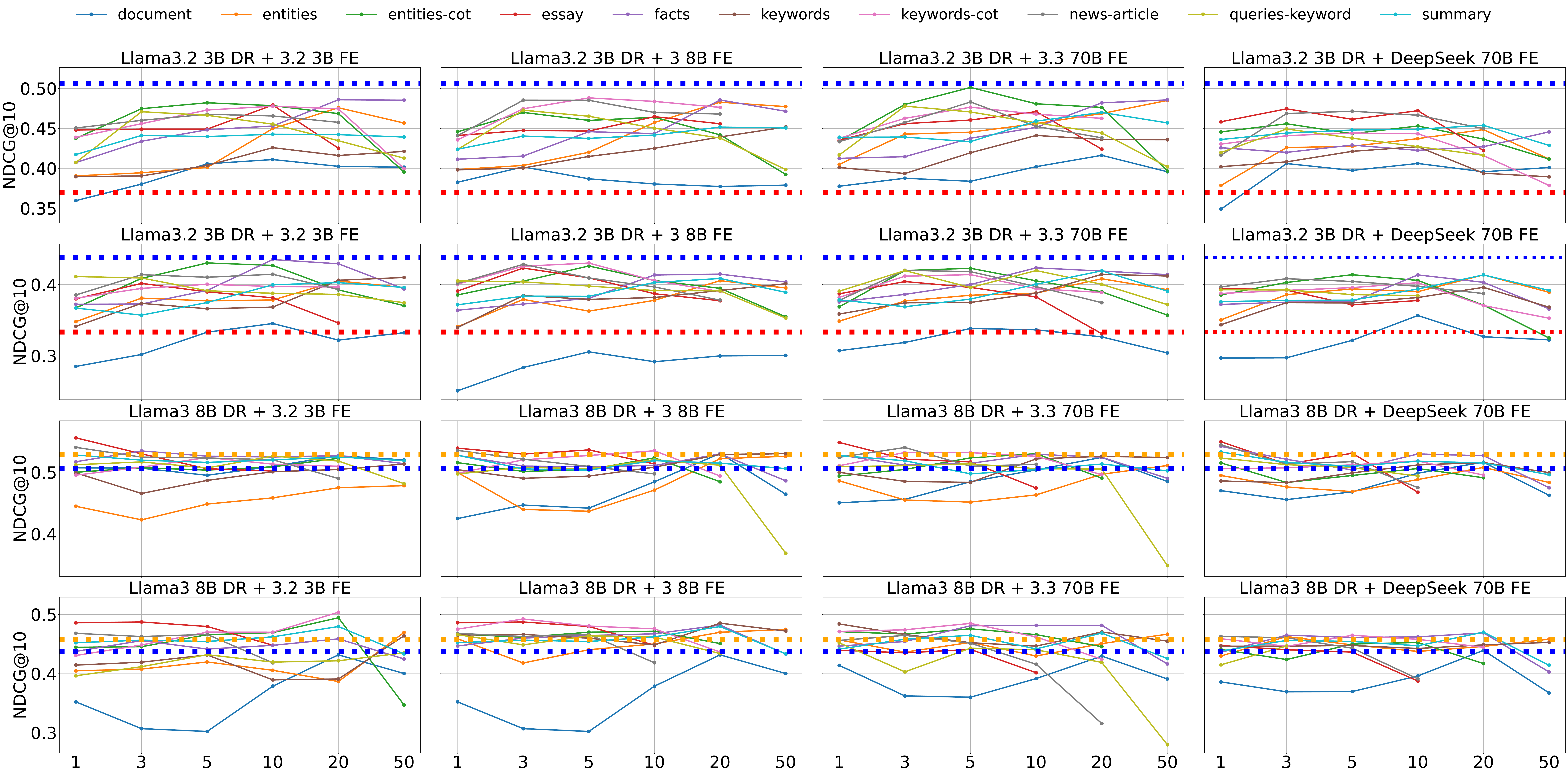}
    \caption{Impact of feature types and PRF depth across dense retrievers of varying scale. Rows 1 and 3 present results for the LLaMA3.2 3B and LLaMA3 8B dense retrievers, respectively, on the TREC DL19 dataset, while rows 2 and 4 show the corresponding results on TREC DL20. Each row varies the feature extractor model and PRF depth applied within the PromptPRF framework. The red, blue, and orange dotted lines indicate baseline performance (no PRF) for LLaMA3.2 3B, LLaMA3 8B, and LLaMA3.3 70B, respectively.}
    \label{fig:features_prf_depth}
\end{figure*}

Figure~\ref{fig:features_prf_depth} demonstrates how the effectiveness of PromptPRF depends on the interplay between PRF depth and feature type. The results show that PromptPRF enables smaller retrievers such as LLaMA3.2 3B to approach the no-PRF baselines of larger retrievers like LLaMA3 8B, and similarly allows LLaMA3 8B with PromptPRF to match or surpass LLaMA3.3 70B. These gains are most observable at moderate PRF depths, and are not limited to features generated by the largest feature extractors, many feature types extracted from smaller LLMs also yield substantial improvements.

% The results emphasize that both PRF depth and feature type play critical roles in determining the effectiveness of PromptPRF. A consistent pattern emerges across models, moderate PRF depths (typically in the range of 3 to 10) yield the highest nDCG@10 scores, while overly deep feedback (e.g., depth 50) often leads to performance degradation. This degradation is particularly observable for verbose or loosely structured features such as Query Keywords and News Articles, which tend to introduce semantic noise and divert the query’s original intent. These findings reinforce prior work suggesting that aggressive query expansion can lead to query drift, especially when the feedback signals are not closely aligned with the query intent.

\noindent\textbf{PRF Depth}: Effectiveness generally peaks at shallow to moderate depths (3–10) across different retrievers and feature extractors. At deeper depths (e.g., 50), performance often drops due to noise accumulation and query drift, particularly when using verbose or loosely structured features. This effect is more noticeable for smaller retrievers, which are less tolerant to excessive feedback.

% The analysis of feature types shows that structured representations, especially Entities-COT and Keywords-COT, consistently enhance performance across PRF depths. These features provide concise, topic-relevant context that is both interpretable and easy to integrate into dense representations. In contrast, factual summarization features such as Facts and Summary provide more stable but less dynamic gains, indicating their utility in delivering neutral, non-distracting augmentations. Full-passage feedback (Document features), while informative at shallow depths, exhibits sharp declines in effectiveness when incorporated at larger depths, again due to noise accumulation.

% \textbf{Feature Extractor Size}: While larger LLMs such as LLaMA3.3 70B and DeepSeek 70B often produce strong features, smaller models like LLaMA3.2 3B can be equally or more effective in certain settings. For example, features from LLaMA3.2 3B yield the best results for the LLaMA3.2 3B dense retriever on DL20, and also outperform larger extractors when paired with LLaMA3 8B on both DL19 and DL20. These results suggest that smaller feature extractors can also offer strong compatibility to different sizes of dense retrievers, the feature extractor effectiveness does not scale with its size.

\noindent\textbf{Feature Type}: Structured and focused features such as Entities-COT and Keywords-COT outperform others across depths and model scales in most cases. They offer semantically rich, compact feedback that integrates well into dense encodings. Summary and Facts provide stable but moderate improvements, indicating a trade-off between informativeness and noise. In contrast, verbose features such as Essay or News Article often yield diminishing or negative returns at higher depths, particularly for large extractors, due to verbosity leads to off-topic drift.

% Comparing the two dense retrievers, LLaMA3 8B outperforms LLaMA3.2 3B across all PRF configurations, as expected given its superior base representation capacity. However, PromptPRF benefits are more observable in LLaMA3.2 3B, particularly at shallow depths (1--5), where feedback compensates for weaker base encodings. In contrast, LLaMA3 8B maintains stable performance even at moderate depths (10--20), reflecting its robustness to broader contextual augmentation. Notably, the relative utility of feature types differs across models, Essay features, for instance, yield greater gains with LLaMA3.2 3B than with LLaMA3 8B, suggesting that smaller models may benefit more from verbose, discursive feedback, whereas larger models derive greater value from concise, structured signals.

\noindent\textbf{Retriever Model Size}: Similar to the observations in Table~\ref{tab:best_ndcg10}, in Figure~\ref{fig:features_prf_depth} we also find that while LLaMA3 8B typically outperforms LLaMA3.2 3B without PRF, both models benefit substantially from PromptPRF. LLaMA3.2 3B sees the largest relative gains, especially at depths 3–10, where feedback compensates for its limited contextual capacity. Notably, LLaMA3 8B with PromptPRF can surpass the no-PRF performance of the much larger LLaMA3.3 70B retriever. These results show that PromptPRF is effective across model scales and can elevate smaller retrievers to match or exceed larger ones.

Overall, these findings show that PromptPRF can boost retrieval effectiveness across model scales. Small dense retrievers can achieve strong gains even with smaller feature extractors, often approaching or surpassing the performance of larger models. While larger feature extractors can help, their benefits are sometimes limited, especially when the feature is verbose or misaligned. For both small and large retrievers, moderate PRF depths and concise, semantically relevant features are keys to maximizing the impact of PromptPRF.

% These findings highlight the need for PRF configurations that are jointly tuned with respect to model capacity, feature type, and feedback depth. For small retrievers, conservative depth settings and richer features are crucial for mitigating drift. For larger retrievers, moderate feedback depths with structured features yields stable, cost-effective gains.

\begin{figure}
    \centering
    \includegraphics[width=\linewidth]{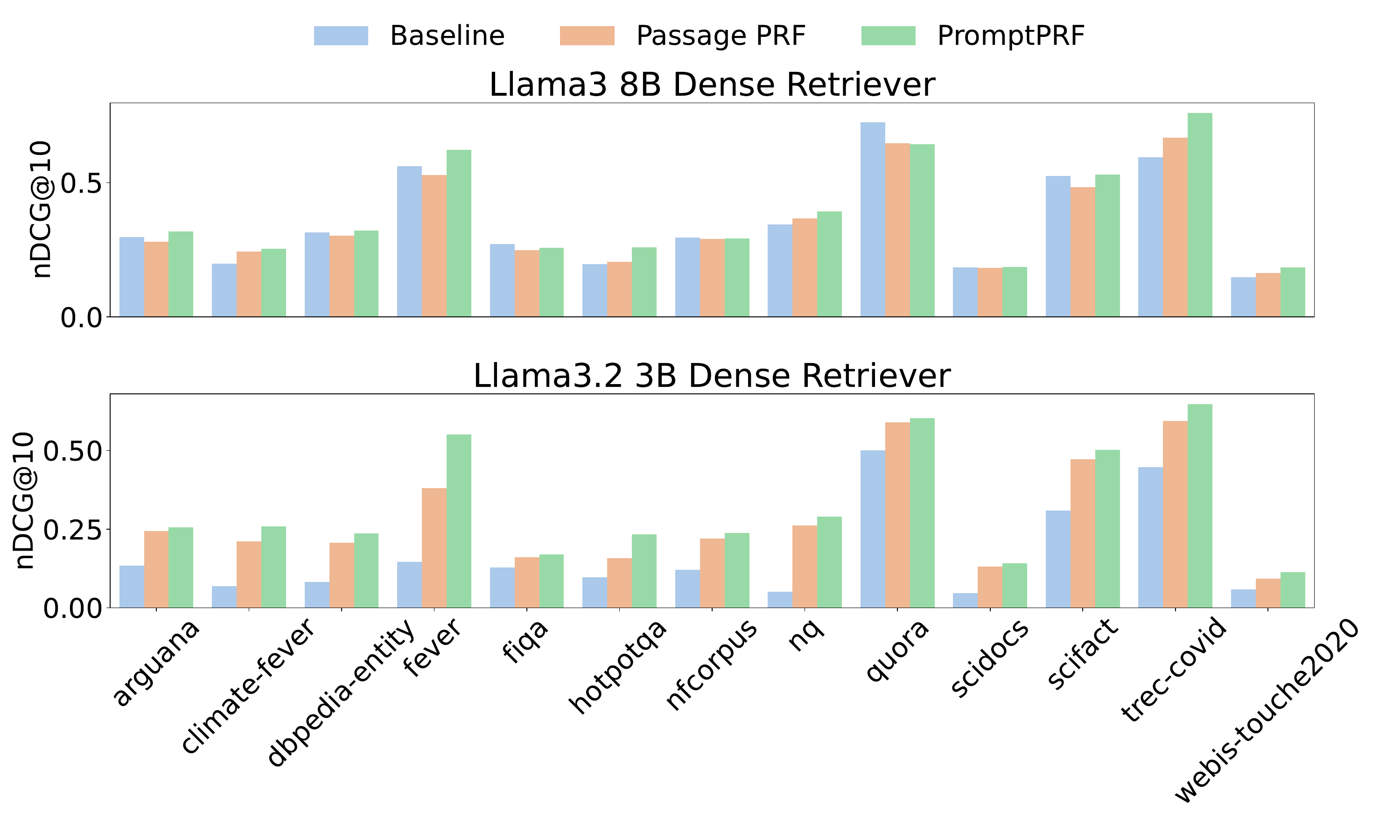}
    \caption{Comparison of nDCG@10 across 13 BEIR datasets using LLaMA3 8B and LLaMA3.2 3B dense retrievers under three configurations: (i) Baseline (no pseudo-relevance feedback), (ii) Passage PRF (using top-k retrieved passage content as feedback), and (iii) PromptPRF (using LLM-generated features as feedback).}
    \label{fig:beir_bars}
\end{figure}

\begin{figure}
    \centering
    \includegraphics[width=\linewidth]{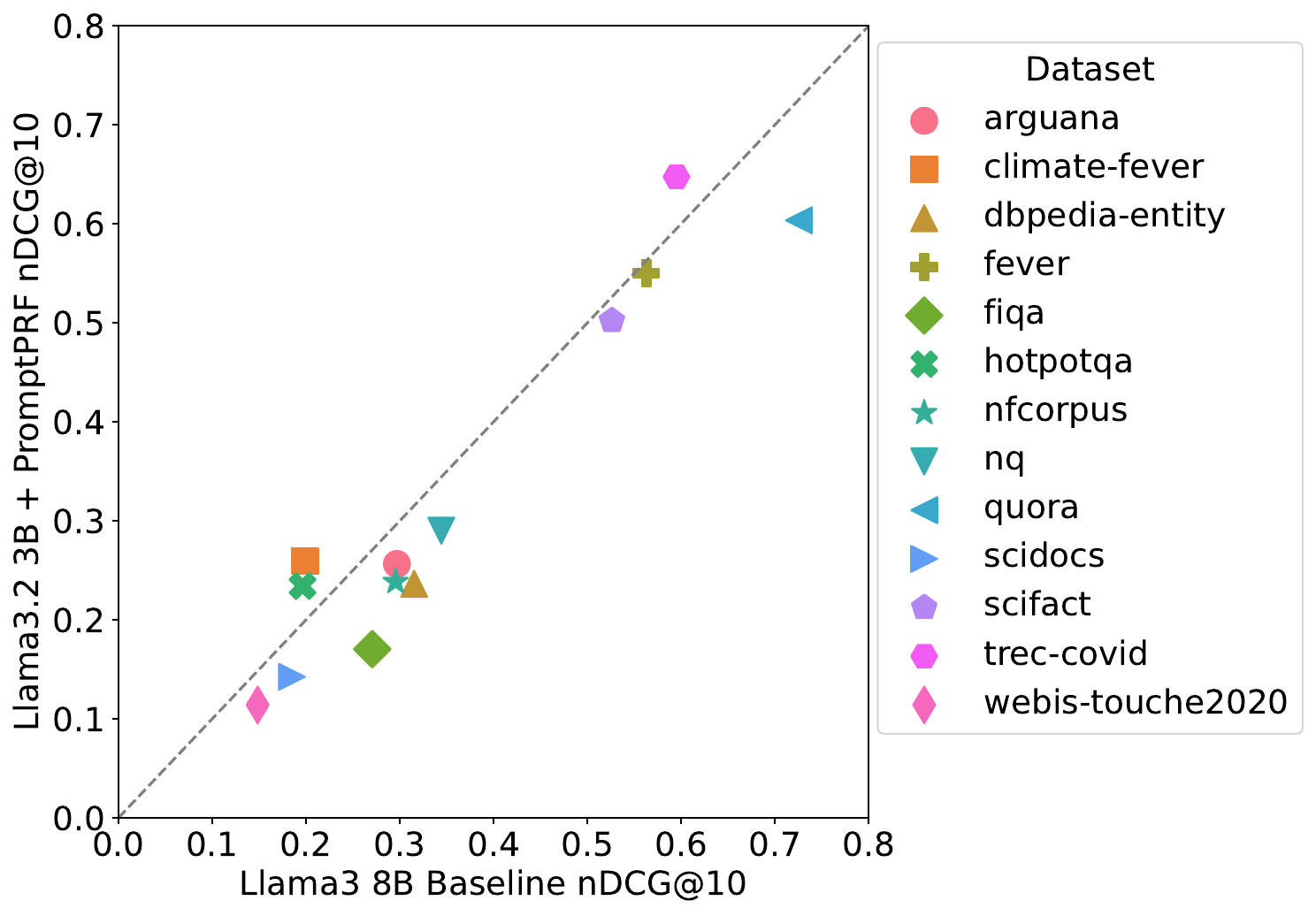}
    \caption{Comparison of retrieval effectiveness between a small LLM-based dense retriever with PromptPRF (LLaMA3.2 3B + PromptPRF) and a larger dense retriever without feedback (LLaMA3 8B Baseline) across BEIR datasets. Each point represents a dataset, with the x-axis showing nDCG@10 for the 8B baseline and the y-axis showing nDCG@10 for the 3B model with PromptPRF. The dashed diagonal line denotes same effectiveness. Points along the line indicate cases where the smaller model with PromptPRF achieves similar effectiveness to the larger baseline model.}
    \label{fig:beir_scatter}
\end{figure}

\subsection{Generalisation to BEIR}

To assess the generalizability of PromptPRF beyond the TREC DL benchmark, we evaluate its performance on the BEIR benchmark, which comprises 13 retrieval tasks spanning diverse domains. As in the previous experiments, we consider two retrievers of different scales, LLaMA3.2 3B and LLaMA3 8B, and compare three configurations for each: (i) Baseline (no PRF), (ii) Passage PRF (direct concatenation of top-$k$ passages), and (iii) PromptPRF (using LLM-generated features extracted from pseudo-relevant passages).

Figure~\ref{fig:beir_bars} presents nDCG@10 scores for all datasets under the three settings. From the figure, PromptPRF consistently outperforms both the Baseline and Passage PRF configurations. The improvements are particularly substantial for LLaMA3.2 3B, where PromptPRF not only boosts performance significantly over the no-PRF and passage-based PRF baselines, but in many cases brings the smaller retriever's effectiveness on par with, or even exceeding, that of LLaMA3 8B without PRF.

This observation is further supported by Figure~\ref{fig:beir_scatter}, which directly compares the performance of LLaMA3.2 3B + PromptPRF against the LLaMA3 8B Baseline. Each point represents a dataset, with its position relative to the diagonal line indicating whether the smaller model with PromptPRF matches or surpasses the larger model without PRF. Most points lie close to or above the diagonal, demostrating that PromptPRF can serve as an equalizer, allowing low-capacity retrievers to approach the performance of much larger models, hence closing the effectiveness gap between small and large retrievers.

These results provide strong empirical support for two key conclusions. First, PromptPRF consistently outperforms naive PRF strategies that use original passage content, validating the value of targeted, LLM-generated features in enhancing query representations. Second, PromptPRF enables small dense retrievers to approach effectiveness levels comparable to much larger models, demonstrating its utility as a cost-effective and deployable alternative in settings where inference efficiency and hardware constraints are critical.

Taken together with our TREC DL findings, the BEIR experiments confirm that structured features, particularly Entities-COT and Keywords-COT, yield the most robust performance improvements. In contrast, feedback based on raw document content or loosely related contextual signals (e.g., News Articles, Query Keywords) is more susceptible to query drift, particularly at larger PRF depths. These insights highlight the importance of adapting PRF design to the model’s capacity and the nature of the dataset, with smaller models benefiting most from conservative, structured, and semantically rich feedback.

In summary, this large-scale evaluation across BEIR validates PromptPRF’s effectiveness as a scalable and retriever-agnostic framework, capable of enhancing both small and large LLM-based dense retrievers in fully zero-shot retrieval settings.

%% file: experiment.tex
\subsection{Experimental Setup} \label{sec_exp_setup}

We study PromptPRF on TREC DL'19~\cite{craswell2020overview}, '20~\cite{craswell2021overview}, and BEIR~\cite{thakur2021beir} benchmarks, using nDCG@10 to assess retrieval effectiveness, as is common practice for these datasets.

%\noindent\textbf{Datasets:} We study PromptPRF on TREC DL'19~\cite{craswell2020overview}, '20~\cite{craswell2021overview}, and BEIR~\cite{thakur2021beir} benchmarks. %, providing a comprehensive evaluation across different retrieval settings.

\begin{figure}[b]
	\includegraphics[width=\columnwidth]{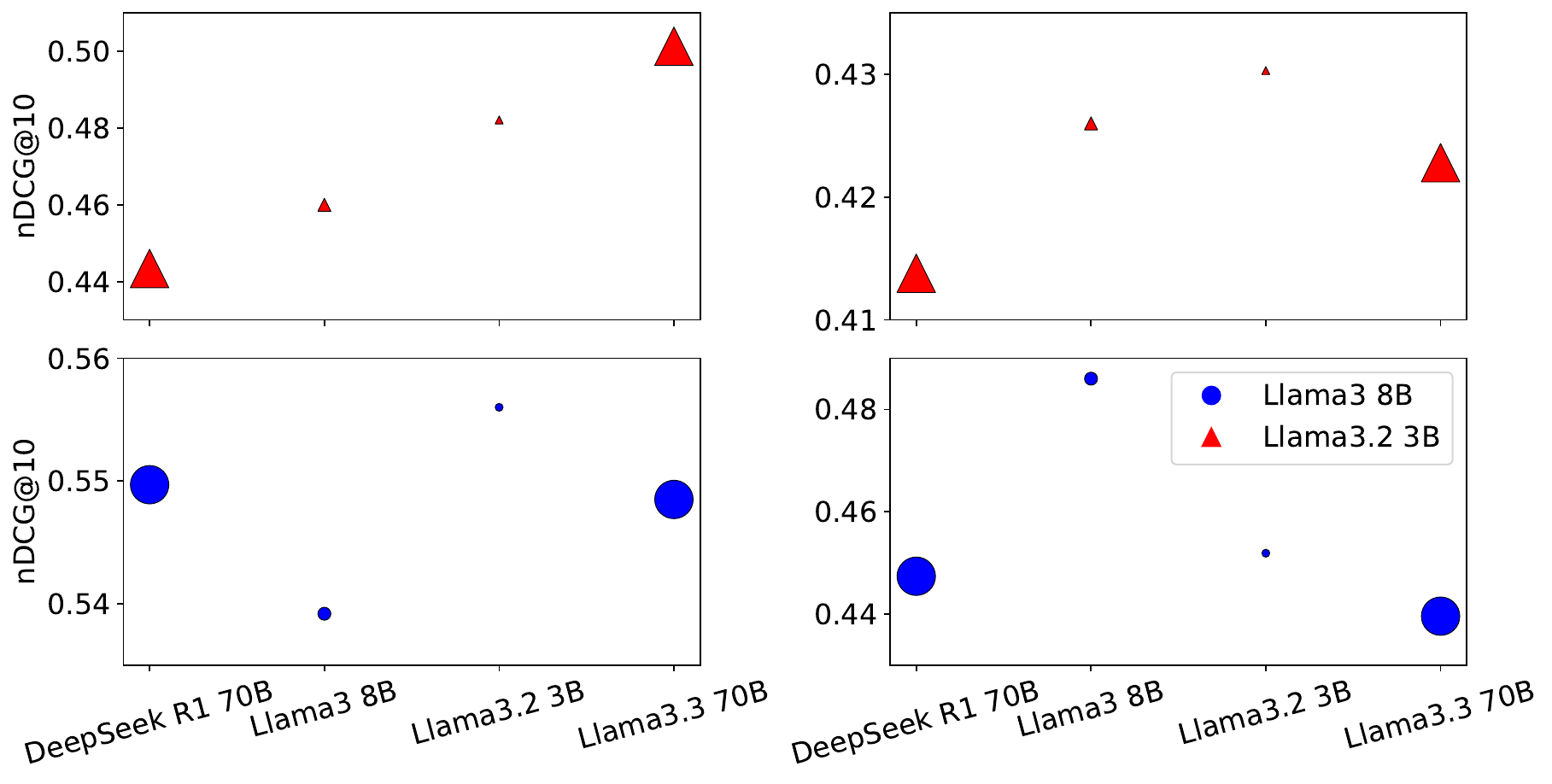}
	\caption{nDCG@10 of PromptPRF when used in combination with a LLaMA3.2 3B based dense retriever (top) and LLaMA3 8B dense retriever (bottom). Results are obtained on DL19 (left) and DL20 (right) and are displayed with respect to different feature extractors (x axis); the marker size represents the size of the feature extractor model (\# parameters).}
	\label{fig:feature_model_effectiveness}
\end{figure}

% \noindent\textbf{Datasets:} We study PromptPRF on both in-domain (TREC DL'19~\cite{craswell2020overview} and '20~\cite{craswell2021overview}) and out-of-domain benchmarks (BEIR~\cite{thakur2021beir}), providing a comprehensive evaluation across different retrieval settings.

%\noindent\textbf{Evaluation Metrics:} To measure retrieval effectiveness, we use nDCG@10 to assess ranking quality, providing insights into precision-oriented performance aspects.

%\noindent\textbf{Rankers:} 
Within the PromptReps framework, we use \textbf{LLaMA3.2 3B Instruct}, \textbf{LLaMA3 8B Instruct}, and \textbf{LLaMA3.3 70B Instruct} as base retrievers without PRF. These models represent dense retrievers of increasing size, allowing us to investigate the effect of model scaling. 
We use the 3B and the 8B backbones for encoding query representations obtained with PromptPRF -- with the focus being to use PRF to augment the 3B dense retriever to close the gap with the 8B dense retriever, and to use PRF to close the gap between the 8B dense retriever and the 70B dense retriever.

%\noindent\textbf{Features Extractors:} 
We use four LLMs for feature extraction: LLaMA3 8B Instruct%\footnote{\label{llama3}https://huggingface.co/meta-llama/Meta-Llama-3-8B-Instruct}~\cite{grattafiori2024llama3herdmodels}
, LLaMA3.2 3B Instruct%\footnote{\label{llama3.2}https://huggingface.co/meta-llama/Llama-3.2-3B-Instruct}~\cite{grattafiori2024llama3herdmodels}
, LLaMA3.3 70B Instruct
%\footnote{https://huggingface.co/meta-llama/Llama-3.3-70B-Instruct}~\cite{grattafiori2024llama3herdmodels}
, and DeepSeek R1 70B.
%\footnote{https://huggingface.co/deepseek-ai/DeepSeek-R1-Distill-Llama-70B}~\cite{deepseekai2025deepseekr1incentivizingreasoningcapability}.
These models vary in parameter size, allowing us to understand how PRF feature quality scales with model size.

%\noindent\textbf{Baselines:} Along with PromptReps without PRF, we include a PRF baseline where the feedback passage(s) text is simply concatenated to the text of the query and then encoded using PromptReps, thus forming a naive PRF method that resembles PromptPRF but does not use feature extractions.

%\noindent\textbf{Hardware \& Code:} 
Experiments are conducted on a server equipped with two H100 GPUs. %, ensuring efficient inference for LLM-based retrieval and feature extraction. 
Code, raw results and prompts are made available at \url{https://anonymous.4open.science/r/PromptPRF-BE44}.

% Need to compress the RQs down to three or two, merge some together.
% \section{Research Questions}
% To systematically evaluate the impact of PRF in LLM-based dense retrieval, we investigate the following research questions:

% \begin{itemize}
%     \item \textbf{RQ1:} Can PRF help smaller LLM-based dense retrieval models achieve performance comparable to larger models?
%     \item \textbf{RQ2:} Does PRF provide further improvements for large LLM-based dense retrieval models?
%     \item \textbf{RQ3:} How do different LLM sizes used for feature extraction impact PRF effectiveness?
%     \item \textbf{RQ4:} What are the effects of different PRF integration strategies on retrieval effectiveness?
% \end{itemize}

% These questions guide our methodology and experimental design, helping us analyze the effectiveness of PRF in different retrieval scenarios.

% \vspace{-6pt}

%% file: ablation.tex
\section{Impact of Rank Information}
%\section{Ablation Study}

%\subsection{Rank-Aware PRF Prompting}

\begin{figure}
    \centering
    \includegraphics[width=\linewidth]{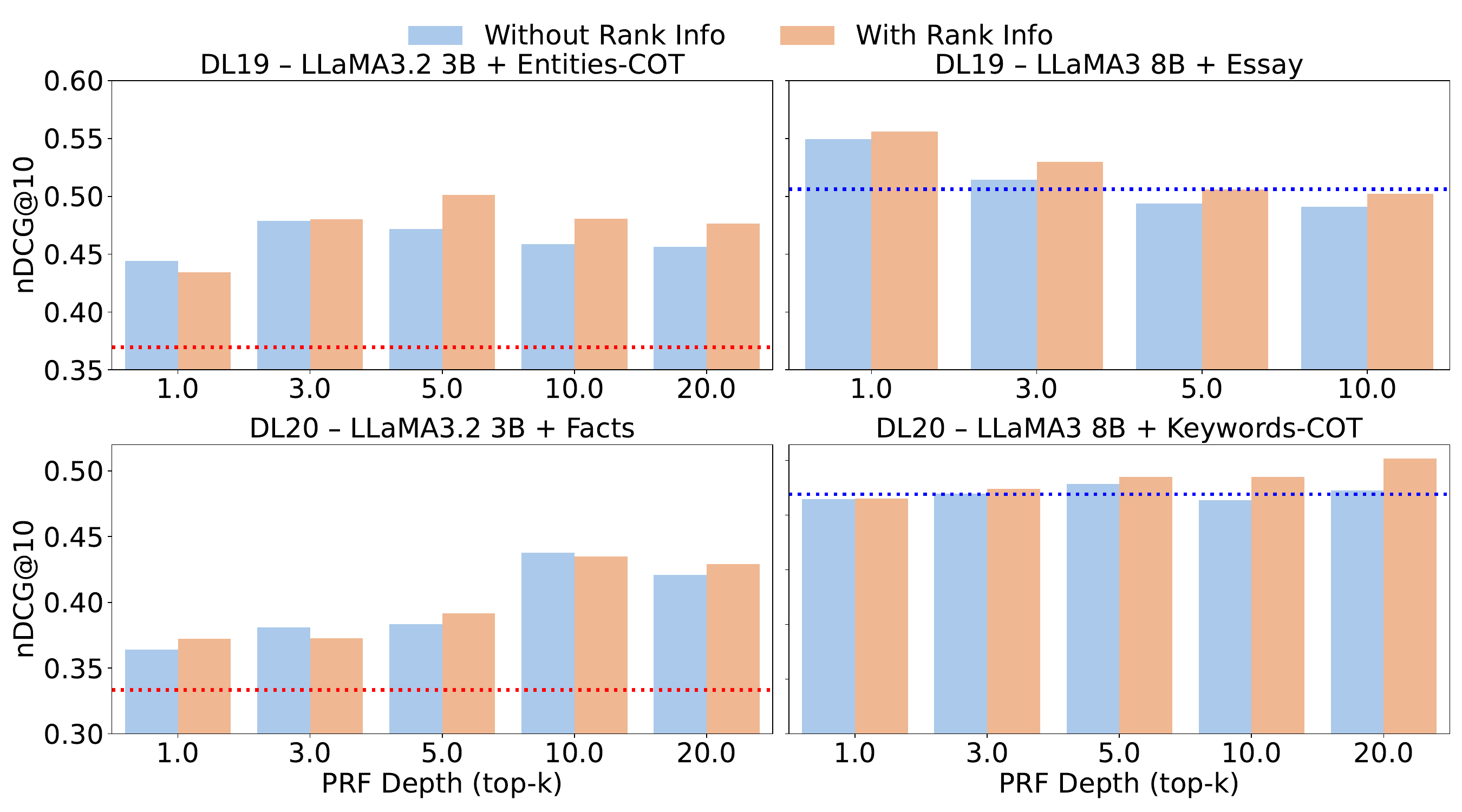}
    \caption{Effect of rank-aware prompting on nDCG@10 across PRF depths for different feature types, retrieval models, and datasets, we choose the best settings for each base dense retriever from Table~\ref{tab:best_ndcg10}. Each subplot compares PromptPRF performance with and without rank information in the prompt. Rank-aware prompting consistently improves effectiveness across configurations, with the largest gains observed at moderate PRF depths. The red dotted line represents LLaMA3.2 3B dense retriever without PRF, blue dotted line represents LLaMA3 8B dense retriever without PRF.}
    \label{fig:rank-aware}
\end{figure}

PromptPRF constructs enhanced query representations by conditioning on feedback signals extracted from top-ranked passages. A key design decision in this process is whether to explicitly encode the retrieval rank of each feedback passage (e.g., "Top 1 Retrieved Passage") within the prompt, or to treat all passages uniformly without positional metadata\footnote{Refer to Appendix~\ref{appendix:rank-ablation} for the specific prompt templates used.}.

%Traditional pseudo-relevance feedback (PRF) methods such as RM3 and Bo1 refine the original query by selecting expansion terms from top-ranked passages and reweighting them using corpus-level statistics. These methods inherently prioritize terms from higher-ranked passages through frequency-based term selection and weighting scheme. In contrast, generative PRF approaches, such as GRF~\cite{Mackie2023gene} and GenPRF~\cite{wang2023generative}, utilize decoder-only architectures that operate on natural language prompts and are decoupled from such statistical signals. Despite this architectural difference, they typically assuming that all feedback is equally informative.

The inclusion on rank information in the feedback signal appears to be unique to PromptReps; previous traditional PRF methods like RM3 and Bo1, or recent generative PRF methods like GRF~\cite{Mackie2023gene} and GenPRF~\cite{wang2023generative} typically assume that all feedback is equally informative and do not include rank information. 
We argue that this assumption overlooks a critical element: rank serves as a proxy for retrieval-time confidence. Higher-ranked passages are more likely to be relevant, and therefore the features extracted from them should have stronger influence during query refinement. PromptPRF addresses this gap by directly encoding rank information into the prompt. %Specifically, in the rank-aware variant, each feedback passage (or extracted feature) is prefixed with its rank position (e.g., "Keywords for Top 1 Retrieved Passage: ..."), allowing the LLM to attend differentially to feedback based on positional information. On the other hand, in the rank-agnostic variant, this metadata is omitted, and features are concatenated without any explicit ranking signal before encoding.
To empirically evaluate the effectiveness of rank-aware prompting, we conduct an ablation study across four diverse configurations, selected from Table~\ref{tab:best_ndcg10} to represent varied model scales, datasets, and feature types (Figure~\ref{fig:rank-aware}): (1) LLaMA3.2 3B retriever on DL19, features: Entities-COT from LLaMA3.3 70B, (2) LLaMA3.2 3B retriever on DL20, features: Facts from LLaMA3.2 3B, (3) LLaMA3 8B retriever on DL19, features: Essay from LLaMA3.2 3B, (4) LLaMA3 8B retriever on DL20, features: Keywords-COT from LLaMA3.2 3B. We report the results in Figure~\ref{fig:rank-aware}.

%\begin{enumerate}
%    \item LLaMA3.2 3B retriever on DL19, features: Entities-COT from LLaMA3.3 70B
%    \item LLaMA3.2 3B retriever on DL20, features: Facts from LLaMA3.2 3B
%    \item LLaMA3 8B retriever on DL19, features: Essay from LLaMA3.2 3B
%    \item LLaMA3 8B retriever on DL20, features: Keywords-COT from LLaMA3.2 3B
%\end{enumerate}

%\subsection{Key Findings}

Across all configurations, we observe that including rank information in the prompt improves retrieval effectiveness, especially at mid-range feedback depths (e.g., $k=5$ or $k=10$). The improvements are not only consistent but also non-trivial, reinforcing that LLMs benefit from structured prompt inputs that reflect the feedback signal's relative strength.

These results demonstrate that by leveraging positional metadata already available in the retrieval pipeline, we can improve prompt-based query refinement without modifying model architecture or requiring training. 
%rank-aware prompting provides a lightweight and effective control mechanism in LLM-based feedback modeling. By leveraging positional metadata already available in the retrieval pipeline, we can improve prompt-based query refinement without modifying model architecture or requiring training. 
The impact is particularly strong for structured and semi-structured features, as rank reinforces their contextual relevance.
We hypothesize that rank-awareness helps in two ways:
\begin{enumerate}
    \item \textbf{Attention biasing}: LLMs may learn to weigh earlier prompt content more heavily, making the order of features critical.
    \item \textbf{Noise mitigation}: Explicit rank may allow the model to better ignore low-quality or off-topic features.
\end{enumerate}

Our findings suggest that rank should be included by default when constructing feedback prompts, especially in zero-shot or inference-only dense retrieval settings.

%% file: cost-analysis.tex
%\section{Cost Analysis and Efficiency-Effectiveness Trade-offs}

\section{Costs and Trade-offs}
Efficient retrieval in real-world applications requires a careful balance between effectiveness and computational cost. In this section, we analyze the infrastructure requirements, query latency, and the trade-offs between retrieval effectiveness and resource efficiency.

%\subsection{Infrastructure Requirements}
%We evaluate the minimum computational resources needed to deploy the different LLMs used in our study. Table~\ref{tab:infra_requirements} presents the disk space, RAM, and GPU requirements for the LLMs used in our experiments as the feature extractors and dense retrievers.

% \inote{refer back to Sec 3}
We have already mentioned the minimum computational resources needed to deploy the different LLMs used in our study (Table~\ref{tab:infra_requirements}).
These specifications highlight that while 3B models can operate on consumer-grade GPUs, larger dense retrievers require high-end hardware with multiple H100 or A100 GPUs. The VRAM overhead of the 70B models further increases deployment complexity, making them significantly more practical for latency-sensitive or resource-constrained deployment scenarios.
% making them significantly more costly.

% Similar observations can be made for the contribution to query latency made by encoding the query (Table~\ref{tab:query_latency}). 
Similar observations apply to the query latency caused by encoding (Table~\ref{tab:query_latency}).
Given that PRF feature extraction is performed offline, the online latency of query encoding is the primary factor  affecting real-time retrieval performance. The dense retriever based on the 3B backbone is approximately 2.7× faster than the 8B one, and 14.5x faster than the 70B, making it significantly more efficient for real-time retrieval applications.
An important advantage of PromptPRF over prior generative feedback methods such as GRF~\cite{mackie2023generative} and Query2Doc~\cite{wang2023query2doc} lies in its query-time efficiency. While PromptPRF may yield lower effectiveness than these approaches, %(Table~\ref{tab:best_ndcg10} and Table~\ref{tab:grf_and_query2doc})
it achieves sizeable improvements in latency and computational requirements, making it significantly more practical for certain real-world deployments.

% In practice, generating features in GRF takes approximately 73 milliseconds per token. For a typical GRF prompt, which produces roughly 2,944 tokens (across 10 distinct features: $64 + 64 + 256 \times 5 + 512 \times 3$), the total latency per query (Only for generation, this does not include query encoding and retrieval) is approximately:

GRF for example relies on on-the-fly feature generation using GPT-3 for every incoming query. Each feature, ranging from entities and facts to summaries, keyword lists, CoT-style reasoning traces, and synthetic passages, is generated from scratch, and is query dependent. This does return effectiveness improvements: For example, BM25+GRF with Essay features obtains an nDCG@10 of 0.6090 on DL19~\cite{GRF}, while a Llama3.2 3B dense retriever with PromptPRF and the same feature type results in an nDCG@10 of 0.5560; and if GRF used all features, then nDCG@10 is pushed to 0.6200. However, for a standard GRF configuration producing ten types of features totaling approximately 2,944 tokens ($64 + 64 + 256 \times 5 + 512 \times 3$), and an average generation time of 73 milliseconds per token, the total latency per query (excluding encoding and retrieval) is $2944 \times 73\ \text{ms} \approx \textbf{219.9 seconds}$. 
%
%\[
%\text{Total latency} = 2944 \times 73\ \text{ms} \approx \textbf{219.9 seconds}
%\]
%
% In addition to the latency bottleneck, the cost of using GPT-3 for this task is substantial. For a typical query of 25 input tokens, the output cost (based on OpenAI's pricing) is roughly 0.044 USD per query. This makes GRF impractical for deployment in real-time or high-throughput retrieval environments.
%
In addition to this latency bottleneck, the cost of using GPT-3 for this task is also substantial. For a typical query of 25 input tokens, the generation of 2,944 output tokens using the GPT-3 API costs approximately 0.044 USD per query (based on OpenAI's pricing), rendering GRF impractical for both low-latency or high-volume applications.

% By contrast, PromptPRF performs offline, query-independent feature generation over the document corpus. These features are generated once and stored, allowing them to be reused across all incoming queries. At query time, no generation is required, enabling zero-shot retrieval with negligible latency and no per-query API cost. This efficiency makes PromptPRF more suitable for real-world and production-scale IR systems.

%In contrast, PromptPRF decouples generation from query execution. It performs offline, query-independent generation of document-derived features, which are stored once and reused across all queries. At inference time, PromptPRF performs retrieval and lightweight prompt construction without any per-query generation, leading to a minimal query-time latency and no API cost on-the-fly. This enables PromptPRF to scale efficiently, supporting deployment in real-time and high-throughput search systems.

Similarly, Query2Doc also offers high effectiveness (e.g., in combination with BM25, it obtains an nDCG@10 of 0.6620 on DL10), but suffers from high latency due to its on-the-fly generation. As reported in the original paper~\cite{wang2023query2doc}, the BM25+Query2Doc setup incurs a total latency exceeding 2,193 milliseconds per query: 16 ms for initial BM25 retrieval, >2,000 ms for pseudo-document generation via a LLM, and 177 ms for the second-stage retrieval.

While PromptPRF does not match the peak effectiveness of GRF or Query2Doc, it offers a favorable effectiveness-efficiency trade-off. This is because PromptPRF decouples feedback generation, which is performed in a query-independent fashion and thus offline, from query execution. The substantial gains in runtime and cost-efficiency make it an attractive choice for production-oriented dense retrieval pipelines, particularly in resource-constrained or latency-sensitive environments.

%% file: case_study.tex
\section{Case Study}

\begin{figure}
  \centering
  \begin{subfigure}[b]{0.235\textwidth}
    \centering
    \includegraphics[width=\linewidth]{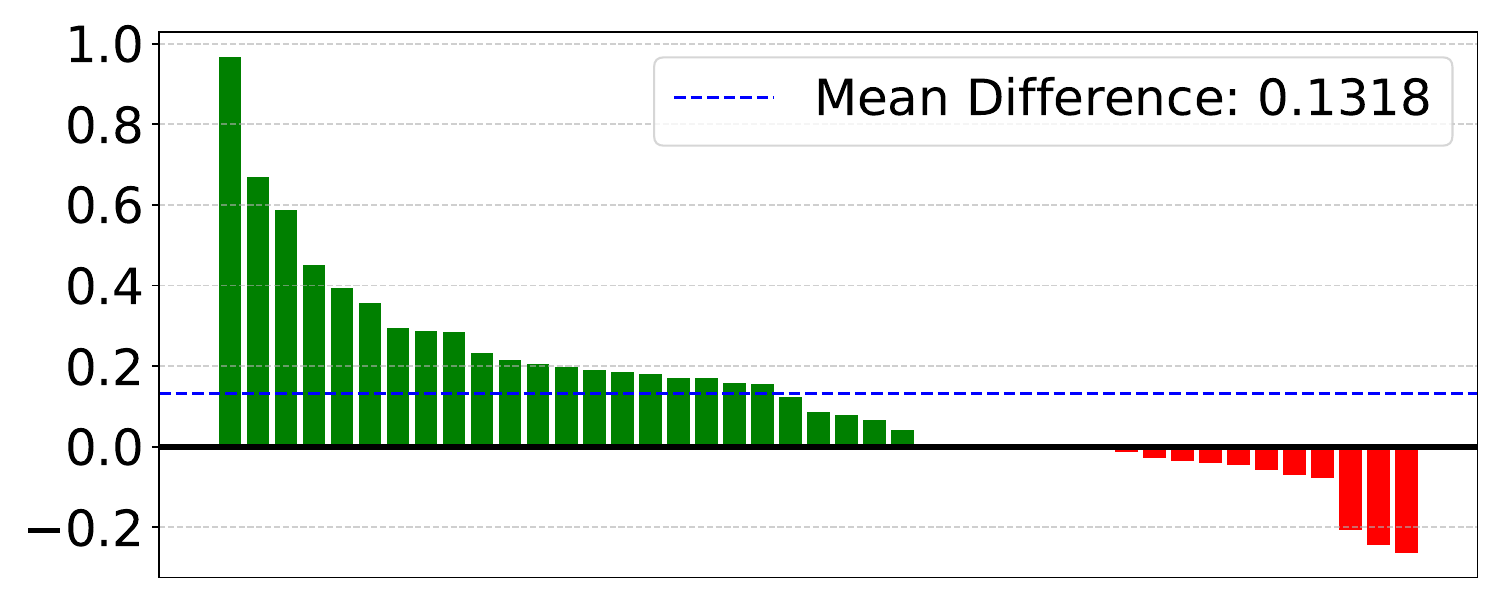}
  \end{subfigure}
  \hfill
  \begin{subfigure}[b]{0.235\textwidth}
    \centering
    \includegraphics[width=\linewidth]{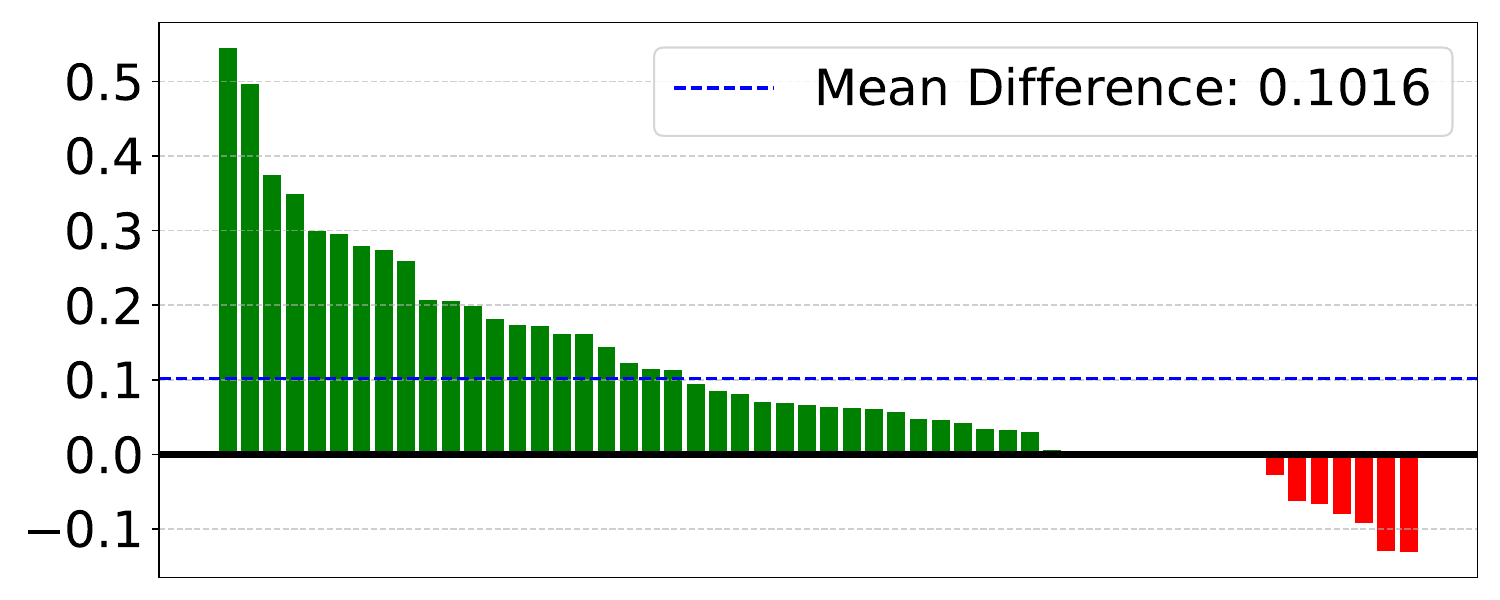}
  \end{subfigure}
  \begin{subfigure}[b]{0.235\textwidth}
    \centering
    \includegraphics[width=\linewidth]{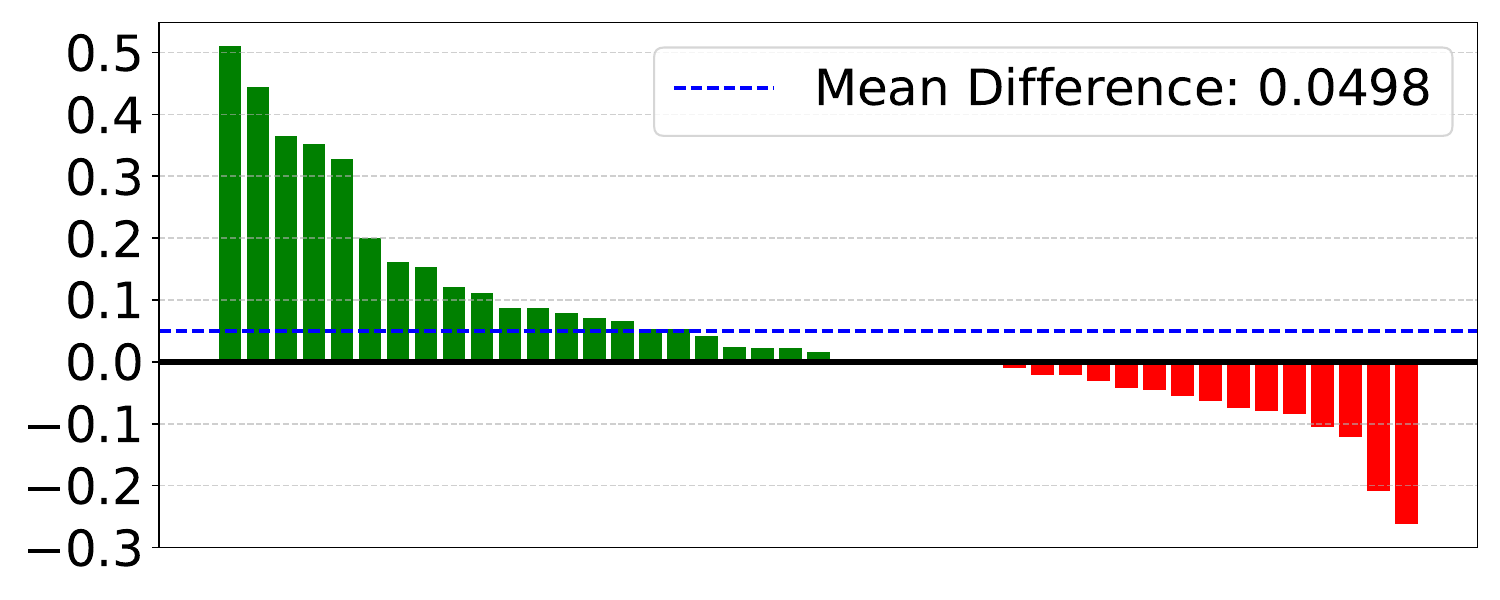}
  \end{subfigure}
  \hfill
  \begin{subfigure}[b]{0.235\textwidth}
    \centering
    \includegraphics[width=\linewidth]{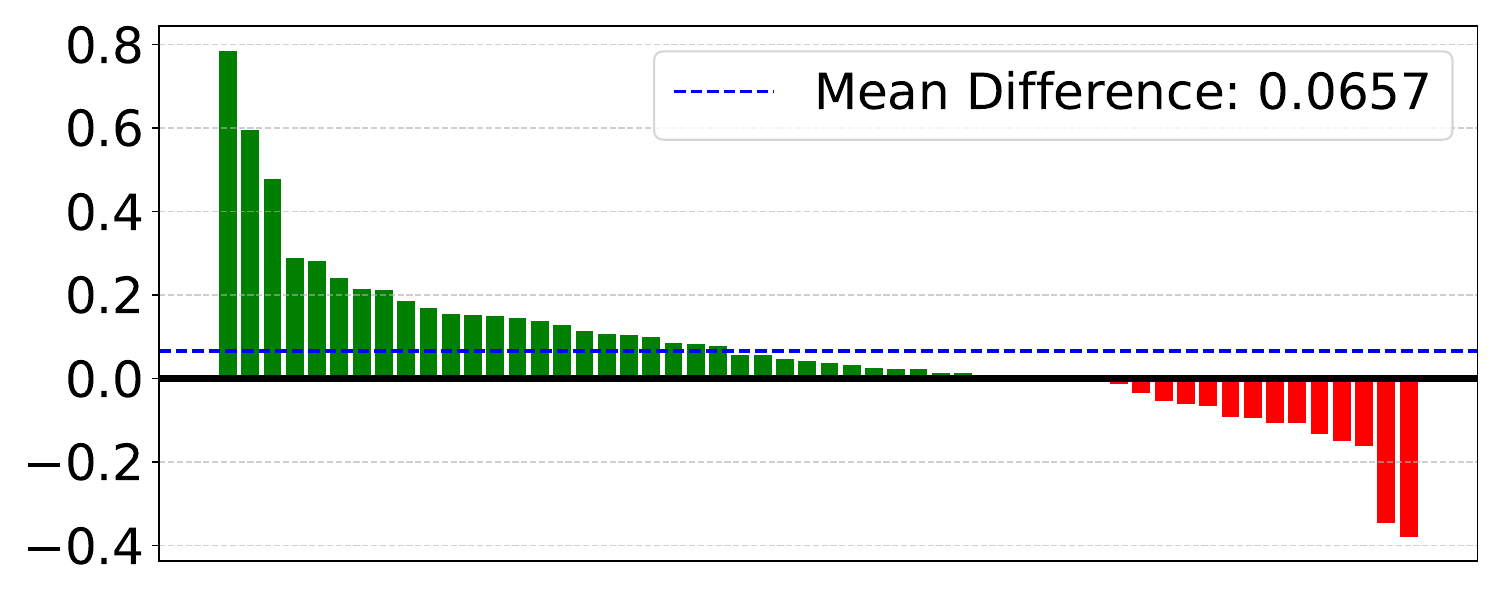}
  \end{subfigure}
% \vspace{-18pt}
  \caption{Per-query analysis of PromptPRF compared to No PRF (PromptReps) for DL19 (left) and DL20 (right) using LLaMA3.2 3B (top) and LLaMA3 8B (bottom) dense retrievers. Each bar represents the difference in nDCG@10 for a single query: Green bars indicate queries for which PromptPRF improves (red bars: degradation). The blue dashed line marks the mean difference. %PromptPRF yields a positive average gain in all settings, with a notable number of large improvements and relatively fewer severe degradations.
  }
  \label{fig:feature_gain_loss}
  % \vspace{-16pt}
\end{figure}

%This section provides a detailed analysis of the effectiveness of PromptPRF, focusing on its comparative performance against No PRF (Initial Retrieval) and Passage PRF (using original passages as feedback). We conduct this analysis using LLaMA3.2 3B and LLaMA3 8B dense retrievers on the DL19 and DL20 datasets, selecting the best-performing feature model for each setting (see Table~\ref{tab:best_ndcg10}). Unlike traditional PRF approaches, PromptPRF is based on pre-generated, query-independent features extracted from top-ranked passages. This design offers both efficiency and modularity but raises questions about how well such features align with specific query intents.

%To examine this, we complement our aggregate evaluation with a query-level gain/loss analysis, exploring where PromptPRF succeeds, where it fails, and why.

Next, we explore where PromptPRF succeeds, where it fails, and why. For this analysis, we use  LLaMA3.2 3B and LLaMA3 8B dense retrievers on the DL19 and DL20 datasets, selecting the best performing feature model for each setting (see Table~\ref{tab:best_ndcg10}).

%\subsection{Aggregate Query-Level Analysis}

Figure~\ref{fig:feature_gain_loss} presents a per-query gain/loss comparison for PromptPRF versus No PRF (i.e. PromptReps). %, sorted by the difference in nDCG@10. Each bar represents the improvement (green) or degradation (red) for a specific query.
We make the following observations:
\begin{enumerate}[leftmargin=14pt,label=\arabic*.,labelsep=6pt, itemsep=0pt, topsep=4pt]
    \item On DL19, PromptPRF improves performance on 22 queries and degrades on 16, with 5 unchanged, yielding a mean nDCG@10 gain of +0.0498 for LLaMA3 8B, and +0.1318 for LLaMA3.2 3B.
    \item On DL20, the improvements are more pronounced: 38 improvements vs. 7 degradations for LLaMA3 8B, and 34 improvements vs. 15 degradations for LLaMA3.2 3B, with mean gains of +0.1016 and +0.0657 respectively.
\end{enumerate}

The long tail of high-magnitude improvements suggests that PromptPRF is particularly effective for queries where the initial retrieval surfaces highly relevant but under-ranked content, allowing the feedback process to steer retrieval more precisely. However, we also observe non-trivial degradations, which manual inspection suggests clustering around ambiguous or broad queries.

%\subsection{Success Cases Overview}

\textbf{Success Examples.} PromptPRF demonstrates strong improvements when top-ranked feedback passages are semantically aligned with the query, and when extracted features (e.g., entities, keywords) preserve the core intent of the original query. For example:

\begin{enumerate}[leftmargin=14pt,label=\arabic*.,labelsep=6pt, itemsep=0pt, topsep=4pt]
    \item \textit{DL19, Query ID: 527433, ``types of dysarthria from cerebral palsy''}. PromptPRF successfully enhanced the relevance of the top-ranked passages by incorporating domain-specific feedback, focusing on medical terms related to cerebral palsy and speech impairments, resulting in a gain of +0.4673.
    \item \textit{DL20, Query ID: 914916, ``what type of tissue are bronchioles''}. PromptPRF effectively captured the biological context of the query, guiding the model toward epithelial tissue-related passages, leading to a significant gain of +0.7853.
\end{enumerate}

These cases reflect the strength of PromptPRF in reinforcing precision, when feedback content is relevant, structured, and topically coherent, the refined query vector becomes both semantically correct and more aligned with relevant documents.

%\subsection{Failure Cases Overview}

\textbf{Failure Examples.} In contrast, degradation cases are often driven by query ambiguity and broad query topics. In these cases, the feedback process causes query drift by broadening the scope of retrieval, leading to the inclusion of top-ranked passages that are less relevant to the query's real intent. For example:

\begin{enumerate}[leftmargin=14pt,label=\arabic*.,labelsep=6pt, itemsep=0pt, topsep=4pt]
    \item \textit{DL19, Query ID: 915593, what is the temperature of hydrogen as a liquid}. % \\
    This query is straightforward in intent but involves complex physical properties. PromptPRF introduced query drift by over-generalising the temperature aspect, leading to the retrieval of passages about the freezing and boiling points of water and other substances rather than focusing specifically on hydrogen’s phase transition temperatures. The drift resulted in a substantial loss of retrieval effectiveness (-0.2617).
    \item \textit{DL20, Query ID: 1122767, what amino produces carnitine}. %\\
    The query seeks a specific biological fact, but PromptPRF expanded the query scope to general discussions on amino acids and protein synthesis. This broadening led to the inclusion of passages on unrelated amino acids and metabolic processes, reducing the effectiveness of the second retrieval (-0.3984).
    \item \textit{DL19, Query ID: 1121402, what can contour plowing reduce}. %\\
    While the query is fairly direct, PromptPRF shifted the focus from the soil erosion aspect to broader topics on agricultural practices and soil properties. This drift caused the model to retrieve documents discussing general farming techniques rather than contour plowing's specific role in erosion control (-0.2907).
\end{enumerate}

These failure cases reflect a common pattern: when the initial retrieval contains broad or tangentially related information, PromptPRF's feedback process, which relies on query-independent features, magnifies this effect rather than narrowing the retrieval focus. %This suggests that PromptPRF may benefit from improved control over semantic drift, particularly for ambiguous or multi-faceted queries. These failure cases also suggest that even high-quality LLM-extracted features can mislead if not contextualized or if the original query is not well represented in the feedback documents.

%\subsection{Failure Due to Passage Type}

\textbf{Failure Due to Passage Type.} While our PromptPRF method demonstrates consistent improvements across a broad range of the BEIR benchmarks, we observe notable effectiveness degradations on specific datasets such as \textsc{Quora} (Figures~\ref{fig:beir_bars} and~\ref{fig:beir_scatter}). %as observable in Figure~\ref{fig:beir_bars} with LLaMA3 8B dense retriever and Figure~\ref{fig:beir_scatter} with \textsc{Quora}. 
A recurring failure mode arises from the nature of the passages themselves, many of which are framed as questions rather than declarative statements. This introduces ambiguity that impairs the utility of these passages as PRF signals. %, particularly when used in conjunction with LLMs.
%
%Consider the following example from the \textsc{Quora} dataset: Passage 4849 is phrased as a question, \textit{"What is a Marketing Director?"}. 
Consider Passage 4849 from \textsc{Quora}: \textit{"What is a Marketing Director?"}. When used as input for feature generation for keywords, even a strong LLM such as LLaMA3.3 70B produces responses that do not engage with the semantic content of the passage, returning instead meta-level replies such as:

\vspace{-2pt}
\renewenvironment{quote}
{\list{}{\leftmargin=0em \rightmargin=0em}%
	\item\relax}
{\endlist}
\begin{quote}
\small
\texttt{"Unfortunately, I don't see a passage provided. Please provide the passage about a Marketing Director, and I'll be happy to generate a bullet-point list of relevant keywords for you!"}
\end{quote}
\vspace{-2pt}

This output reflects a failure to ground the generation in the content of the actual passage and is symptomatic of the model interpreting the input as an incomplete prompt rather than a stand-alone knowledge source. %Consequently, the generated response lacks any relevant or informative signals that could enhance the query representation. 
When incorporated as part of the PRF mechanism, such responses effectively introduce noise, leading to severe query drift and substantial drops in retrieval effectiveness.

This failure highlights a broader limitation: LLM-based PRF methods that rely on generative rewriting or feature extraction from raw passages are sensitive to the syntactic and pragmatic framing of those passages. In cases where the corpus predominantly contains interrogative or conversational text (e.g., \textsc{Quora}), the model may misinterpret the task context, yielding invalid feedback material. This finding emphasises the need for more robust input conditioning strategies or filtering mechanisms that can effectively detect and mitigate the use of ill-formed or ambiguous passages.

%\subsection{Summary}
%
%Our query-level analysis reveals that PromptPRF is highly effective for a substantial subset of queries, particularly those where the initial retrieval surfaces passages that are topically coherent and semantically aligned with the query intent. In such cases, PromptPRF can extract or generate high-quality features that substantially improve relevance ranking, often recovering relevant documents missed by baseline dense retrievers.
%
%Nonetheless, PromptPRF remains vulnerable to semantic drift when applied to ambiguous queries, overly broad feedback scopes, or corpora containing noisy or uninformative passages. These failure cases showcase the importance of aligning feedback signals more tightly with the query's underlying intent. Future work may address these limitations by incorporating query-aware filtering mechanisms, adaptive confidence thresholds for generated features, or hybrid feedback strategies that condition LLM generation on both the query and its retrieval context to mitigate drift and improve robustness.

%% file: conclusion.tex
\section{Discussion and Conclusion}
\label{conclusion}
% In this paper, we devised PromptPRF, a generative PRF method for LLM-based dense retrieval that improves effectiveness without incurring high query-time costs.

In this paper, we devised PromptPRF, a generative PRF method for LLM-based dense retrieval, PromptPRF, that, without imposing large costs at query time (query latency, compute, and hardware requirements), provides an alternative to backbone scaling-up for improving retrieval effectiveness which might be well suited to certain real-world applications, including on resource-constrained hardware and strictly constrained query-latency tasks.
For example, our results show that the effectiveness of a 3B dense retriever can be improved to become close (and in some cases indistinguishable) to those of an 8B dense retriever, departing from conventional expectations based on LLM scaling laws~\cite{kaplan2020scaling, hoffmann2022training}, which predict performance gains primarily through increasing model size and training data~\cite{fang2024scaling}.

The flexibility and efficiency of PromptPRF open up several possibilities for future research. First, the contribution of individual feedback passages or features could be further refined by learning or tuning specific weights $\beta_i$, which were fixed to $\beta_i=1$ in this work. Second, while our analysis confirms the utility of rank-aware prompting, the precise impact of encoding rank information in the prompt deserves further investigation. Finally, future studies could explore the compositional effects of combining multiple feature types, including hybrid strategies that balance interpretability, brevity, and semantic richness.